\documentclass[twocolumn,showpacs,preprintnumbers,amsmath,amssymb]{revtex4}

\usepackage{epsfig,psfrag}
\usepackage{dcolumn}
\usepackage{bm}
\usepackage{graphicx}
\usepackage{color}

\begin{document}
\title{Theory of non-equilibrium transport in the SU(N) Kondo regime}
\author{Christophe Mora$^1$}
\author{Pavel Vitushinsky$^2$}
\author{Xavier Leyronas$^3$}
\author{Aashish A. Clerk$^2$}
\author{Karyn Le Hur$^4$}
\affiliation{$^1$Laboratoire Pierre Aigrain, ENS, Universit\'e Denis Diderot 7, CNRS; 24 rue Lhomond, 75005 Paris, France}
\affiliation{$^2$Department of Physics, McGill University, Montr\'{e}al,
 Qu\'{e}bec, Canada, H3A 2T8}
\affiliation{$^3$Laboratoire de Physique Statistique de l'Ecole Normale Sup\'erieure
associ\'e au CNRS et aux Universit\'es Paris 6 et Paris 7,
24 rue Lhomond, F-75005 Paris, France}
\affiliation{$^4$Department of Physics, Yale University, New Haven,
 Connecticut,  USA, 06520}


\begin{abstract}
Using a Fermi liquid approach, we provide a comprehensive treatment of the current and current noise through a quantum dot
whose low-energy behaviour corresponds to an SU($N$) Kondo model, focusing on the case $N=4$ relevant to carbon nanotube dots.  We show that for general $N$,
one needs to consider the effects of higher-order Fermi liquid corrections even to describe low-voltage current and noise.  We also show that the noise exhibits complex behaviour
due to the interplay between coherent shot noise, and noise arising from interaction-induced scattering events.  We also treat various imperfections 
relevant to experiments, such as the effects of asymmetric dot-lead couplings. 
\end{abstract}

\pacs{71.10.Ay, 71.27.+a, 72.15.Qm}

\maketitle

\section{Introduction}

The Kondo effect has long served as a paradigm in the field of strongly correlated electron physics.  It is perhaps the simplest example of a system where many-body interactions can give rise to highly non-trivial behavior: its essence involves nothing more than a localized magnetic impurity which is exchange coupled to conduction electrons in a metal.  Despite having been studied for over 40 years, interest in Kondo physics shows no sign of abating.  A large part of this continued interest has been fueled by recent advances allowing the controllable realization of unusual Kondo effects in nanostructures.  These include multi-channel Kondo effects \cite{nozieres1980}, where there are many conserved flavours of conduction electrons:  such systems can give rise to non-Fermi liquid physics, and have recently been realized using semiconductor quantum dots \cite{potok2007}.  Another class of exotic Kondo effects are so called SU($N$) Kondo effects, where $N>2$.  Such systems involve only a single channel of conduction electrons, but the effective spin of the impurity and conduction electrons is greater than $1/2$.  While such systems are still described at low-energies by a Fermi liquid fixed point, the properties of this Fermi liquid are modified in several interesting ways compared to the spin-$1/2$ case \cite{mora2009b}.  The case $N=4$ has received particular attention due to its realizability in double \cite{borda2003,lehur2004,lopez2005} and triple quantum dots \cite{numata2009} as well as carbon nanotube quantum dots
\cite{choi2005, jarillo2005, choi2006, finkelstein2007, lehur2007}. 

Research on Kondo physics has also been spurred by the possibility of studying experimentally its behaviour when driven out-of-equilibrium, where non-equilibrium is either achieved by the application of a drain-source voltage across a quantum dot \cite{silvano2002,gg2008}, or by externally radiating a quantum dot \cite{elzerman2000}.  The non-equilibrium induced by a voltage has been the subject of a number of recent theoretical works \cite{rosch2003,kehrein2005,mehta2006,doyon2007,boulat2008,anders2008,lehur2008}.

In this paper, we will focus on a topic which combines two of the above avenues of Kondo research:  we will study non-equilibrium charge transport through a 
voltage-biased quantum dot exhibiting an SU($N$) Kondo effect, focusing on the low-temperature regime where the physics is described by an effective Fermi liquid theory.  We present calculations for both the non-linear conductance as well as for the current noise.  As has been stressed in a number of recent papers \cite{sela2006, mora2008, vitu2008}, the fluctuations of current through a Kondo quantum dot are extremely sensitive to the two-particle interactions associated with the underlying Fermi liquid theory.  This was first discussed in the case of the standard SU($2$) Kondo effect by 
Sela et al.~\cite{sela2006} , and was even measured for this system in a recent experiment by Zarchin et al.~\cite{heiblum2008}.    As discussed in Refs.~\cite{vitu2008,mora2008}, the situation becomes even more interesting for $N>2$, as now one must deal with the interplay between coherent partition noise (due to the zero-energy transmission coefficient through the dot not being one) and the interaction-induced scattering events.  Of particular interest is the case $N=4$, which can be realized in carbon nanotube quantum dots.  Very recently, current noise in such a system has been measured experimentally by Delattre et al. \cite{delattre2009}, though not in the low-temperature Fermi liquid regime we describe here.  

The results presented here both clarify and extend those presented in Refs.~\cite{mora2008,vitu2008}, as well as provide details underlying the calculational approach.  Particular attention is given to the role of higher-order Fermi liquid corrections, something that was not correctly treated in previous works (see erratum, Ref.~\cite{mora2009}).  We show clearly how in the $N=4$ case, such corrections lead to an effective shift
of the Kondo resonance with applied bias voltage.  As a result, the non-linear conductance does not increase with voltage, as would be expected from a simple picture of the Kondo resonance
as a resonant level sitting above the Fermi energy.  These Fermi-liquid energy shifts are absent in the usual $N=2$ Kondo effect.  We also describe the experimentally-relevant case where
there is an asymmetry in the coupling between the quantum dot and the source and drain electrodes.  Such an asymmetry has not been investigated thoroughly in previous works. 

The remainder of this paper is structured as follows.  In Sec.~\ref{sec-model}, we outline the basic description of our model and the Fermi liquid approach.  Sec.~\ref{sec-current} and~\ref{currentnoise} are devoted to providing a detailed discussion of our results for both the conductance and the shot noise, as well as details on their derivation. In Sec.~\ref{results}, we summarize our main results for the conductance and shot noise of a SU(N) Kondo quantum dot, and conclude.

\section{Model description}\label{sec-model}
\subsection{Kondo Hamiltonian}

We give here a compact synopsis of the quantum dot model we study, and how it gives rise to Kondo physics.
The dot connected to the leads is described by the following 
Anderson Hamiltonian~\cite{Krishna1980}
\begin{equation}\label{anderson}
\begin{split}
H & =  H_{\rm D} + H_{\rm L} + H_{\rm T} = \varepsilon_{\rm d} 
\sum_\sigma n_{\sigma}  + 
 U \sum_{\sigma<\sigma'} n_{\sigma} n_{\sigma'} \\[1mm]
&+ \sum_{k,\sigma} \varepsilon_k ( c_{{\rm L},k\sigma}^\dagger
c_{{\rm L},k\sigma} + c_{{\rm R},k\sigma}^\dagger
c_{{\rm R},k\sigma} )  \\[1mm]
&+ \sum_{k,\sigma} (t_{\rm L} c_{{\rm L},k\sigma}^\dagger 
+ t_{\rm R} c_{{\rm R},k\sigma}^\dagger) d_{\sigma} + h.c.
\end{split}
\end{equation}
$c_{{\rm L/R},k\sigma}$ is the annihilation operator for an electron
of spin $\sigma=1 \ldots N$ and energy $\varepsilon_k = \hbar v_F k$ (measured from
the Fermi energy $\varepsilon_F$)
confined on the left/right lead. $d_\sigma$ is the electron operator of the dot and 
$n_\sigma = d^\dagger_\sigma d_\sigma$ the corresponding density. 
$U$ denotes the charging energy, $\varepsilon_{\rm d}$ the single particle energy on the dot and
 $t_{\rm L/R}$ the tunneling matrix elements from the dot to
the left/right lead.
The general case of asymmetric leads contacts is parametrized by 
$t_{\rm L} = t \cos \theta$, $t_{\rm R} = t \sin \theta$ with
$\theta = [ 0 , \pi/2 ]$. $\theta = \pi/4$ recovers the symmetric case.
The rotation in the basis of leads electrons
\begin{equation}\label{rota}
\begin{pmatrix} b_{k\sigma} \\ a_{k\sigma} \end{pmatrix}
= \begin{pmatrix} \cos \theta & \sin \theta \\ \sin \theta & - \cos \theta 
  \end{pmatrix}
\begin{pmatrix} c_{{\rm L},k\sigma} \\ c_{{\rm R}, k\sigma} \end{pmatrix},
\end{equation}
decouples the $a_{k\sigma}$ operators from the dot variables.
The Kondo screening then  involves only the $b_{k\sigma}$ variables.
In the symmetric case, $\theta = \pi/4$, $b_{k\sigma}$ and $a_{k\sigma}$
represent respectively even and odd wavefunctions with the dot placed at $x=0$.

We consider in this work the Kondo limit where the charging energy $U$ is
by far the largest energy scale. Below this energy, the charge degree of 
freedom on the dot is quenched to an integer value and does not fluctuate.
For $\varepsilon_{\rm d} = U  (1-m-m/N)$, the number of electrons is 
$\sum_\sigma n_\sigma = m$. The virtual occupation
of other charge states by exchange tunneling with the leads is accounted for
by the standard Schrieffer-Wolff transformation~\cite{schrieffer1966} (or second order
perturbation theory). It transforms Eq.~\eqref{anderson}
to the Kondo Hamiltonian $H = \sum_{k,\sigma} \varepsilon_k ( b_{k\sigma}^\dagger
b_{k\sigma} + a_{k\sigma}^\dagger a_{k\sigma} ) + H_K$  where
\begin{equation}\label{kondo1}
H_K = J_K \sum_{k\sigma,k',\sigma'} \left( d^\dagger_\sigma d_{\sigma'} - \frac{m}{N} 
\delta_{\sigma,\sigma'} \right) b_{k'\sigma'}^\dagger b_{k\sigma},
\end{equation}
and $J_K = \frac{t^2}{U} \frac{N^2}{m (N-m)}$. This Hamiltonian acts in the subspace
constrained by  $\sum_\sigma n_{\sigma} = m$.
In this paper, we concentrate on the choice $\varepsilon_{\rm d} = U  (1-m-m/N)$
for which potential scattering terms vanish after the Schrieffer-Wolff transformation.
Including potential scattering in the formalism is possible, for example along the line
of Ref.~\cite{affleck1993}. It however remains outside the scope of this work where we focus on
the asymmetric dot-lead couplings.

The lead electrons transform under the fundamental representation of SU(N).
With exactly $m$ electrons, the localized spin on the dot transforms as a representation
of SU(N) corresponding to a single column Young tableau of $m$ boxes.
A basis of generators for this SU(N) representation is formed by the $N^2-1$ traceless
components $S^{\sigma,\sigma'} = d_{\sigma}^\dagger d_{\sigma'} - (m/N) \delta_{\sigma,\sigma'}$
with $(\sigma,\sigma')\ne (N,N)$. This basis can be used~\cite{parcollet1998} to 
rewrite Eq.~\eqref{kondo1} as an antiferromagnetic coupling 
\begin{equation}\label{kondo2}
H_K = J_K  \vec{S} \cdot \vec{T},
\end{equation}
between the impurity (dot) spin $\vec{S} = (S^A, A=1,\ldots,N^2-1)$ and the  spin
operator of the lead electrons taken at $x=0$, $\vec{T} = ( \sum_{k,k',\sigma,\sigma'}
b_{k\sigma}^\dagger t^A_{\sigma,\sigma'} b_{k'\sigma'},A=1,\ldots,N^2-1)$.
The $N \times N$ matrices $t^A$ are generators of the  fundamental representation of SU(N),
while $S^A$ are $\frac{N !}{m ! \, (N-m)!} \times \frac{N !}{m ! \, (N-m)!}$ matrices 
acting on states with $m$ electrons.

Starting from high energies, $J_K$ grows under renormalization. It presages
the complete screening of the dot spin by the formation of a many-body 
SU(N) singlet in the ground state.
A large body of studies has shown that the strong coupling fixed point
that dominates at low energy is a Fermi liquid one.
Exact results from the Bethe-Ansatz~\cite{bazhanov2003} find low energy exponents 
that characterize a Fermi liquid. Writing the Kondo Hamiltonian~\eqref{kondo1}
in terms of current, Affleck~\cite{affleck1990} has shown by completing the square
that the impurity spin can be absorbed by lead electrons. The resulting
(conformal field) theory is that of free fermions and it is believed 
to describe the strong coupling fixed point. It shows a simple translation of energies 
in the spectrum corresponding to an electron phase shift imposed
by the Friedel sum rule
\begin{equation}\label{zerops}
\delta_0 = \frac{m \, \pi}{N}.
\end{equation}
The identification of the leading irrelevant operator at this fixed point
yields Fermi liquid behavior~\cite{affleck1993}.
Alternatively and following Ref.~\cite{nozieres1980}
 the ground state of Eq.~\eqref{kondo2} has been shown~\cite{parcollet1998} to be
a singlet state. Turning the coupling to the leads does not destabilize
this singlet leading again to Fermi liquid exponents.
Finally, Numerical Renormalization Group (NRG) calculations have confirmed
this picture for SU(2)~\cite{wilson1975} and 
SU(4)~\cite{borda2003,lehur2004,choi2005,choi2006}.

\subsection{Fermi liquid theory}

We now discuss in detail the Fermi liquid theory for the Kondo effect, 
first introduced  by Nozi\`eres~\cite{nozieres1974a+nozieres1978}. 
It describes the low energy regime - the vicinity of
the strong coupling fixed point - and allows one to make quantitative predictions
even in an out-of-equilibrium situation.
In Ref.~\cite{mora2009b}, the Fermi liquid theory of Nozi\`eres has been extended
with the introduction of the next-to-leading order corrections to the strong coupling 
fixed point. These corrections are necessary in the SU(N) case for observables like
the current and the noise since their energy ($k_B T$, $e V$ or 
$\mu_B B$) dependence is  mostly quadratic.

The Kondo many-body singlet (also called the `Kondo cloud') having been formed, we wish to describe how lead electrons scatter off it.
At low energies, two channels open: an elastic and an inelastic one.
Both take place at the dot position $x=0$.
Elastic scattering is described by an energy-dependent phase shift.
At the Fermi level $\varepsilon_F$, it is equal to $\delta_0$, see Eq.~\eqref{zerops}.
We expand the phase shift around the Fermi energy,
\begin{equation}\label{pshift}
\delta_{\rm el} (\varepsilon) = \delta_0 + \frac{\alpha_1}{T_K} \varepsilon 
+ \frac{\alpha_2}{T_K^2} \varepsilon^2,
\end{equation}
where the energy $\varepsilon$ is measured from  $\varepsilon_F$.
$\alpha_1$ and $\alpha_2$ are dimensionless coefficients of order one.

It is instructive to think of the {\it elastic} scattering off the Kondo singlet in terms of an effective non-interacting resonant level model (RLM), where this effective
resonance represents the many-body Kondo resonance.  This is the picture of the Kondo effect provided by slave-boson mean-field theory~\cite{coleman1984}, and is an exact description of the SU($N$) Kondo effect in the large $N$ limit \cite{newns1987}.  Note that for finite $N$, one must also deal with true two-particle scattering off the singlet, something that will never be captured by the RLM; we thus only use it to obtain insight into the elastic scattering properties.  
In the RLM picture, the first two terms in the phase shift in Eq.~(\ref{pshift}) are attributed to a  Lorentzian scattering resonance
centered at $\varepsilon_K = T_K \cot \delta_0$ with a width $\propto T_K$ \cite{vitu2008}.  In the SU($4$) case, one thus finds that the Kondo resonance is centered at a distance $\varepsilon_K = T_K$ above the Fermi energy, giving a heuristic explanation for the fact that the low-energy transmission coefficient through the dot is only $T=1/2$.  The fact that the Kondo resonance sits above the Fermi energy is indeed seen in exact NRG calculations of the impurity spectral density \cite{choi2005,choi2006}.

The low energy expansion of the RLM phase shift $\delta ( \varepsilon ) = \arctan \left( 
\frac{T_K}{\varepsilon_K - \varepsilon} \right)$ also gives the form Eq.~\eqref{pshift} with
$\alpha_2/\alpha_1^2 = \cot \delta_0$.  Note that there is no apriori reason that this relation must hold
for the expansion of the true phase shift, as the correspondence to a non-interacting resonant level is not exact.   Despite this caveat,
one finds that the prediction from the RLM picture is quite good even at a quantitative level.   
The exact relation between $\alpha_1$ and $\alpha_2$ is extracted~\cite{mora2009b} from the Bethe ansatz solution~\cite{bazhanov2003} and reads
\begin{equation}\label{FLs3}
\frac{\alpha_2}{\alpha_1^2} = \frac{N-2}{N-1} \, \frac{\Gamma(1/N)\tan(\pi/N) }
{\sqrt{\pi}\Gamma\left(\frac{1}{2} + \frac{1}{N} \right)} \cot \delta_0,
\end{equation}
where $\delta_0$ is given Eq.~\eqref{zerops}. 
In the SU(2) case, or more generally for a half-filled
dot with $m=N/2$, $\alpha_2= 0 $, corresponding to a Kondo resonance centered at the Fermi level.
This is expected for a model where particle-hole symmetry is not broken.
In the SU(4) case, Eq.~\eqref{FLs3} gives~\cite{mora2009} $\alpha_2/\alpha_1^2  \simeq 1.11284$ instead of
$1$ in the RLM. As expected, the agreement becomes even better at larger $N$, and RLM result is indeed 
the $N=\infty$ limit of Eq.~\eqref{FLs3}.

The phase shift in Eq.~\eqref{pshift} completely characterizes the low-energy elastic scattering off the Kondo singlet.  For further calculations, it
is useful to describe it using a Hamiltonian formulation.   The free Hamiltonian describing purely elastic scattering is given by
\begin{equation}\label{elastic}
H_0 = \sum_{k,\sigma} \varepsilon_k ( b_{k\sigma}^\dagger
b_{k\sigma} + a_{k\sigma}^\dagger a_{k\sigma} ).
\end{equation}
Decoupled from the outset, the $a_{k\sigma}$ variables are the same as in the 
original model. 
 In contrast, the $b_{k\sigma}$ variables have been modified to now include the elastic
phase shift $\delta_{\sigma}(\varepsilon)$ in Eq.~\eqref{pshift}.
This point will be expanded on in Sec.~\ref{sec-current} when we discuss the calculation of the 
current.

We turn now to inelastic effects, which arise from quasiparticle interactions in the Fermi liquid theory.
These interactions can be written in a Hamiltonian form~\cite{mora2009b}
\begin{equation}\label{interaction}
\begin{split}
& H_{\rm int} = \frac{\phi_1}{\pi \nu^2 T_K} \sum_{\sigma < \sigma',\{ k_i \}} 
: b_{\sigma,k_1}^\dagger b_{\sigma,k_2}
b_{\sigma',k_3}^\dagger b_{\sigma',k_4}  : \\[2mm]
& + \frac{\phi_2}{4 \pi \nu^2 T_K^2} \sum_{\sigma < \sigma',\{ k_i \}}
( \sum_i \varepsilon_{k_i}) 
: b_{\sigma,k_1}^\dagger b_{\sigma,k_2}
b_{\sigma',k_3}^\dagger b_{\sigma',k_4}  :  \\[2mm]
& - \frac{\chi_2}{\pi \nu^3 T_K^2}  \sum_{\sigma < \sigma'<\sigma'' \atop
\{ k_i \}} \! \! \!
: b_{\sigma,k_1}^\dagger b_{\sigma,k_2}
b_{\sigma',k_3}^\dagger b_{\sigma',k_4} b_{\sigma'',k_5}^\dagger b_{\sigma'',k_6}  :,
\end{split}
\end{equation}
where  $:\,\,:$ denotes normal ordering and 
$\nu = 1/ (h v_F)$ is the density of state for 1D fermions moving along one direction.
To summarize, the Fermi liquid theory is generated by the Hamiltonian $H  = H_0 + H_{\rm int}$,
given by Eqs.~\eqref{elastic} and~\eqref{interaction}, with the elastic phase shift~\eqref{pshift}.
In fact, Eqs.~\eqref{pshift} and~\eqref{interaction} correspond to a systematic expansion of
the energy~\cite{nozieres1974a+nozieres1978,mora2009b}, compatible with the SU(N) symmetry and 
the Pauli principle. It includes all first and second order terms in the low energy 
coupling strength $\propto 1/T_K$.

The great advantage of the Fermi liquid approach is that it can also be applied to non-equilibrium
situations.
Note that the Fermi level $\varepsilon_F$ appears twice in the above equations:  it defines the  reference for 
energies in the elastic phase shift~\eqref{pshift} and also for the normal ordering in Eq.~\eqref{interaction}.
When the system is put out-of-equilibrium, for instance when each lead has its own 
Fermi level, $\varepsilon_F$ loses its meaning as a Fermi level and becomes merely
an {\it absolute} energy reference. This can be used to relate~\cite{mora2009b} the coefficients 
$(\alpha_1, \alpha_2, \phi_1, \phi_2, \chi_2)$ as we shall show below.

\subsection{{\it Kondo floating} and perturbation theory}

\begin{figure}
\includegraphics[width=5.5cm]{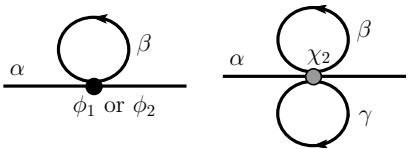}
\caption{\label{loops} Examples of Hartree diagrams for the self-energy built 
from Eq.~\eqref{interaction}. The full dots (resp. black and grey)
indicate vertices with four or six external
lines. $\alpha$, $\beta$ and $\gamma$ denote spins.}
\end{figure}

To make progress in calculating physical observables at low energies, 
we will treat the interacting part $H_{\rm int}$ (c.f. Eq.~\eqref{interaction}) of the Fermi-liquid Hamiltonian perturbatively.
Among the various diagrams built from Eq.~\eqref{interaction}, it is convenient to
separate the trivial Hartree contributions to the electron self-energy
from the more complicated diagrams.
The former are obtained by keeping an incoming and an outgoing line and by closing all
other external lines to form loops as shown Fig.~\ref{loops}. 
The resulting diagrams
are then in correspondence with the diagrams describing scattering by a local potential.
Therefore they can be included in the elastic phase shift,
\begin{equation}\label{phase2}
\begin{split}
& \delta_{\sigma} (\varepsilon)  = \delta_0 + \frac{\alpha_1}{T_K} \varepsilon 
+ \frac{\alpha_2}{T_K^2} \varepsilon^2 - \sum_{\sigma'\ne\sigma}
\Bigg( \frac{\phi_1}{T_K} \delta N_{0,\sigma'}
 + \frac{\phi_2}{2 T_K^2}  \\[1mm] &  ( \varepsilon \, \delta N_{0,\sigma'}
+ \delta E_{1,\sigma'}  )
- \frac{\chi_2}{T_K^2} \sum^{\sigma'/\sigma''\ne \sigma}_{\sigma''<\sigma'} \delta N_{0,\sigma'}
\delta N_{0,\sigma''} \Bigg),
\end{split}
\end{equation}
where we have defined $\delta N_{0,\sigma} = 
\int d \varepsilon  \delta n_{\sigma} (\varepsilon)$
and $\delta E_{1,\sigma} = 
\int d \varepsilon \, \varepsilon \delta n_{\sigma} (\varepsilon) $.
$\delta n_{\sigma} (\varepsilon) = n_{\sigma} (\varepsilon) - \theta(\varepsilon_F - \varepsilon)$
is the actual quasiparticle distribution 
($n_{\sigma} (\varepsilon_k) = \langle  b_{k\sigma}^\dagger b_{k\sigma}  \rangle$)
relative to the ground state with Fermi energy $\varepsilon_F$.
We see again that $\varepsilon_F$ sets the reference in Eq.~\eqref{phase2} for both $\varepsilon$
and $\delta n_{\sigma} (\varepsilon)$.
Including Hartree diagrams is essentially tantamount to a mean-field treatment of the interaction
term Eq.~\eqref{interaction}. On a physical level, these Hartree terms can be interpreted as a mean-field energy shift of the Kondo resonance arising from a finite quasiparticle population and
their interactions.
We shall see that in the case of an SU($4$) Kondo quantum dot, these Hartree terms play a significant role in determining the 
non-linear conductance; this is not the case in the more conventional SU($2$) Kondo effect.

While the idea of perturbatively treating $H_{\rm int}$ is straightforward enough, a possible weakness of the Fermi liquid approach is the number of seemingly undetermined parameters in Eqs.~\eqref{pshift}
and~\eqref{interaction}.  The standard Fermi liquid treatment of the Kondo effect allows one to relate the coefficients $\alpha_1$ and $\phi_1$ via the so-called `floating' of the Kondo resonance (to be discussed below); these coefficients correspond to leading-order Fermi liquid corrections.  However, for transport quantities in the general SU($N$) Kondo case, we will see that the remaining coefficients, corresponding to higher-order corrections, are also important.  Luckily, these too can be related to one another using a novel and powerful extension of the Kondo {\it floating} recently proposed in  Ref.~\cite{mora2009b}.
%
It allows to relate the different phenomenological coefficients of Eqs.~\eqref{pshift}
and~\eqref{interaction}; we describe the basic reasoning involved in what follows.

The Kondo resonance is a many-body phenomenon that results from
the sharpness of the Fermi sea boundary~\cite{kondo1964}. Physically, conduction
electrons build their own resonance. The structure of this resonance
is therefore changing with the conduction electron occupation numbers, as
Eq.~\eqref{phase2} shows explicitly.
However it can not depend on $\varepsilon_F$, which is a {\it fixed} 
energy reference. This idea is implemented by shifting the Fermi
level $\varepsilon_F$ by $\delta \varepsilon_F$ while keeping the {\it absolute} energy 
$\varepsilon_F+ \varepsilon$
and the {\it absolute} occupation numbers $n_\sigma (\varepsilon)$ fixed in Eq.~\eqref{phase2}. 
As a result $\varepsilon \to \varepsilon - \delta \varepsilon_F$ and 
$\delta n_\sigma (\varepsilon) \to \delta n_\sigma (\varepsilon) + \theta(\varepsilon - \delta
\varepsilon_F) - \theta (\varepsilon)$. Imposing
invariance of the phase shift leads to the following Fermi liquid identities
\begin{subequations}\label{FLs}
\begin{align}
\label{FLs1}
\alpha_1 &= (N-1) \phi_1, \\[1mm]\label{FLs2}
\alpha_2 = \frac{N-1}{4} \, \phi_2, & \qquad \phi_2 = (N-2) \, \chi_2,
\end{align}
\end{subequations}
where the first relation~\eqref{FLs1} was initially derived for the general SU($N$) case
by Nozi\`eres and Blandin~\cite{nozieres1980}.  

Note that an alternative way to derive Eqs.~\eqref{FLs} is to insist that the {\it entire structure} of the Kondo resonance simply translates in energy when we dope the system with quasiparticles in a way that corresponds to a simple increase of the Fermi energy~\cite{mora2009b}.  Nozi\`eres' original derivation of Eq.\eqref{FLs1} in the SU($2$) case \cite{nozieres1974a+nozieres1978} also used this idea, but restricted attention to an initial state with no quasiparticles.  Eqs.\eqref{FLs2} follow when we apply the same reasoning to an initial state having some finite number of quasiparticles.  Note that for SU(2), or a half-filled dot ($m=N/2$),
$\alpha_2 =0$ from Eq.~\eqref{FLs3} so that $\phi_2 = 0$ and $\chi_2=0$. The next-to-leading order corrections
all vanish in agreement with previous works on the ordinary SU(2) 
case~\cite{nozieres1974a+nozieres1978,meir2002,golub2006,sela2006}.    

It is worth mentioning that the second generation of Fermi liquid terms $(\alpha_2,\phi_2,\chi_2)$ can also be derived in the framework
of conformal field theory. In Ref.~\cite{mora2009b}, a single cubic Casimir operator is given, which reproduces the three terms
corresponding to the coefficients $\alpha_2,\phi_2$, and $\chi_2$. The identities~\eqref{FLs2} are then automatically satisfied.

\begin{figure}
\includegraphics[width=7.5cm]{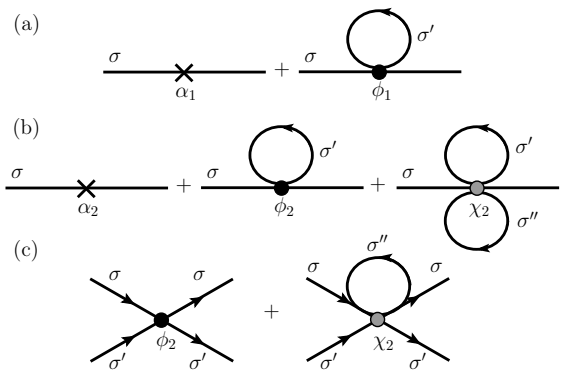}
\caption{\label{gauge} Diagrammatic construction for the independence of observables
in $\varepsilon_F$. Crosses correspond to elastic scattering. Two- and three-particle
interactions are represented by, respectively, black and grey full circles.
Many diagrams in the perturbative expansion in $H_{\rm int}$~\eqref{interaction}
exhibit a dependence in $\varepsilon_F$. Nevertheless, it is possible to gather and combine
those diagrams to produce $\varepsilon_F$-invariant forms. The combination (a) that appears
in the irreducible self-energy does not depend on $\varepsilon_F$ as a result of Eq.~\eqref{FLs1}.
Combination (b) is a second invariant, thanks to
Eqs.~\eqref{FLs2}, contributing to the irreducible self-energy. 
(a) and (b) together imply the phase shift invariance discussed in the text.
Finally, the four-particle vertices of (c) can always be combined to cancel the 
dependence in $\varepsilon_F$ thanks to Eq.~\eqref{FLs2}.
 Apart from (a), (b) and (c), all other diagrammatic parts involve
energy differences in which the reference $\varepsilon_F$ naturally disappears.
The combinations (a), (b) and (c) can be understood as emerging from Ward identities
related to the U(1) gauge symmetry. For example, Eq.~\eqref{FLs1} has been 
shown~\cite{hewson1993,hewson1994} to derive
from a Ward identity with a vanishing charge susceptibility.}
\end{figure}

The {\it floating} of the Kondo resonance (and resulting conditions)
also has an important consequence for calculations of observables in the presence of a voltage:  the results will
not depend on where one decided to place the dot Fermi energy $\varepsilon_F$ within the energy window defined by the chemical potentials of the leads.  On a technical level, this is because, by virtue of Eqs.~\eqref{FLs}, any shift $\delta \varepsilon_F$ of the dot Fermi energy will be completely compensated by a corresponding shift in the Hartree contributions arising from the quasiparticle interactions.  This invariance is explained in detail in Fig.~\ref{gauge}.  Note also that this invariance has physical consequences as well:  it implies, for example, that the current is not affected by the capacitive coupling to the leads (in the Kondo limit). 

Given the above invariance, 
it is convenient for calculations to choose the Fermi level such
that 
\begin{equation}\label{choche}
\delta N_{0,\sigma} = \int d \varepsilon   \delta n_\sigma (\varepsilon) =  0,
\end{equation}
so that any closed fermionic loop built from an energy-independent
vertex vanishes. For this choice of position, $\delta N_{0,\sigma}$ vanishes
which greatly simplifies the phase shift expression~\eqref{phase2}.
Moreover, the $\chi_2$ vertex in Eq.~\eqref{interaction} does not contribute to
the current and the noise when the perturbative calculation is stopped at second
order. The reason for that is that the  $\chi_2$ vertex is already second order and
can only appear once. Its six legs are connected to at most two current vertices
so that at least two of these legs must connect to form a closed loop
implying a vanishing contribution.
In contrast to these simplifications, $\delta E_{1,\sigma}$ in Eq.~\eqref{phase2}
remains generally different from zero due to the energy 
dependence of the $\phi_2$ vertex in Eq.~\eqref{interaction}.

On may wonder whether the physical argument of the {\it floating} of the Kondo
resonance, as presented in Ref.~\cite{mora2009b} and repeated in this paper,
is sufficient to extend the results of this paper to higher orders Fermi liquid corrections.
Applying the {\it floating} argument to the next (third) order, one obtains an incomplete set
of relations between the coefficients such that some of them remain undetermined.
In the language of conformal field theory, it means that more than one operator is involved
at each (higher) order. How to relate the coefficients of those operators is a rather difficult problem. In the SU(2) case, a solution
was given by Lesage and Saleur~\cite{lesage1999a+lesage1999b}.

We finally turn to the discussion of the Fermi liquid model renormalization.
Treated naively, the model leads to divergences in physical quantities. 
It is regularized~\cite{affleck1993} by introducing an energy cutoff $D$ 
(different from the original band
width of the model) larger than typical energies of the problem but
smaller than $T_K$. Energies in Eq.\eqref{elastic} are therefore restricted to 
the window $[-D,D]$. The dependence of observables in $D$ is then 
removed by adding counterterms in the Hamiltonian. It is strictly equivalent
to the introduction of cutoff $D$ dependence 
in the  coupling constants~\cite{mora2003} ($\alpha_i$, $\phi_i$, etc). The corresponding
counterterms are discussed in Appendix~\ref{counterterms}.

\section{Current calculation}\label{sec-current}

We now outline the calculation of the current using the Fermi liquid theory described in previous sections.  
Again, the complete Hamiltonian is  $H = H_0 + H_{\rm int}$ (c.f.~Eqs.~(\ref{elastic},\ref{interaction})), corresponding to respectively
to elastic and inelastic scattering; the approach will be to treat  $H_{\rm int}$ as a perturbation. 
Slightly abusing terminology, we will include all Hartree contributions arising in perturbation theory in the free Hamiltonian
$H_0$; $H_0$ will thus correspond to the elastic phase shift given in Eq.~\eqref{phase2}.
Contributions to the current which only involve $H_0$ (thus defined) will be referred to as the `elastic current'.  
$H_{\rm int}$ is then added perturbatively, without Hartree diagrams, in order
to compute the corrections due to inelastic scattering.

\subsection{The current operator}\label{subsec-current}

The current operator at $x$ is generally given by
\begin{equation}
\hat{I} (x) = \frac{e \hbar}{2 m i} \sum_\sigma \left( 
\psi_\sigma^\dagger (x) \partial_x  \psi_\sigma (x) 
- \partial_x \psi_\sigma^\dagger (x) \psi_\sigma (x) \right)
\end{equation}
where $m$ is the electron mass. Various expressions can be obtained for the current
depending on which basis it is expanded. It is convenient~\cite{mora2008} 
in our case to choose
the basis of scattering states that includes completely elastic (and Hartree terms)
scattering, {\it i.e.} the phase shift~\eqref{phase2}, 
and that correspond to eigenstates of the single-particle scattering matrix.  Such states will
have waves incident from both the left and right leads.  This is in contrast to another standard choice \cite{vitu2008}, which is to use scattering
states which {\it either} have an incident wave from the left lead, or from the right.  We refer to such states as the `left/right' states.

We first discuss our scattering states in first quantization.
Eigenfunctions corresponding to the $a_{k\sigma}$ variables do not see the dot or
the Kondo effect. Using Eq.~\eqref{rota}, they read
\begin{equation}\label{eig1}
\psi_{a,k} (x) = \begin{cases} \sin \theta ( e^{i (k_F+k) x} - e^{-i (k_F+k) x} ) 
 \quad x<0,
\\ \cos \theta ( e^{i (k_F+k) x} - e^{-i (k_F+k) x} )  \quad x>0,
		 \end{cases}
\end{equation}
where $\theta - \pi/4$ measures the asymmetry of the coupling to the leads, 
see Eq.~\eqref{rota},
and the eigenenergies $\hbar v_F k$ are measured from the Fermi level $\varepsilon_F$. 
The situation is more
complicated for the $b_{k\sigma}$ variables. The associated eigenfunctions at small $x$,
close to the dot, depend on the complex ground state wavefunction of the Kondo problem.
They are not known, and in fact it can not even be reduced to a one-particle problem.
 However, we can write the eigenfunctions
far from the dot,
\begin{equation}\label{eig2}
\psi_{b,k} (x) = \begin{cases} \cos \theta ( e^{i (k_F+k) x} - {\cal S}_k 
e^{-i (k_F+k) x} )  \quad x<0,
\\ \sin \theta ( e^{-i (k_F+k) x} - {\cal S}_k e^{i (k_F+k) x} )  \quad x>0,
		 \end{cases}
\end{equation}
where the ${\cal S}$ matrix is related to the phase shift~\eqref{phase2},
${\cal S}_k = e^{2 i \delta(\varepsilon_k)}$ at eigenenergy 
$\varepsilon_k = \hbar v_F k$.
The eigenstates~\eqref{eig1} and~\eqref{eig2} have the same energy.
They can be combined to give the left and right scattering states
with the energy-dependent transmission $T (\varepsilon) = \sin^2(2 \theta)
\sin^2 (\delta(\varepsilon))$. In the SU(2)  case (or generally particle-hole
symmetric case), $\delta_0 = \pi/2$ and 
the system is closed to unitarity for symmetric leads coupling.

We come back to second quantization and project the electron operator
$\psi_\sigma (x)$ over the eigenstates~\eqref{eig1} and~\eqref{eig2}. 
Conservation of the current implies that $\hat{I} (x)$ does
not depend on $x$. We choose an arbitrary $x<0$ far from the dot,
$\hat{I}_L$ is the current at $x$ and $\hat{I}_R$ at $-x$.
If $\hat{I}$ denotes the conserved current, $\hat{I} =  \hat{I}_L
= \hat{I}_R$. The combination 
$\sin^2 \theta  \, \hat{I}_L + \cos^2 \theta \, \hat{I}_R$ leads to
the compact expression
\begin{equation}\label{current}
\begin{split}
\hat{I} & = \frac{e}{2 \nu h} \, \sum_\sigma \Big( \sin 2 \theta \,
[ a^\dagger_\sigma (x) b_\sigma(x) - a_\sigma^\dagger(-x) {\cal S} b_\sigma(-x) + {\rm h.c.} ] \\[2mm]
& - 2 \cos 2 \theta \, [ a_\sigma^\dagger(x) a_\sigma(x) - a_\sigma^\dagger(-x) a_\sigma(-x) ] \Big), 
\end{split}
\end{equation}
with $b_\sigma(x) = \sum_k b_{k\sigma} e^{i k x}$ and ${\cal S} b_\sigma(x)
= \sum_k {\cal S}_k b_{k\sigma} e^{i k x}$. 
Physically, operators taken at $x$ ($-x$)
correspond to incoming (outgoing) states~\cite{blanter2000}.
The second line in Eq.~\eqref{current}
turns out not to contribute to the mean current, the noise or any 
moment of the current.

Before proceeding with the calculation, it is worth noting that in the SU(2) case, 
the proximity to the unitary situation allows 
a simpler treatment~\cite{kaminski2000}. The current is written $\hat{I}  = I_{\rm u} 
- \hat{I}_{\rm BS}$ with $I_u = \frac{2 e^2}{h} V$. All quantum or
thermal fluctuations are included in the backscattering current $\hat{I}_{\rm BS}$
which can be written in terms of $a_{k\sigma}$ and $b_{k\sigma}$ 
operators~\cite{golub2006,gogolin2006b}. 
 However, the range of application of this approach is restricted 
 to the SU(2) case with a completely symmetric leads coupling. In any other
situations neglecting fluctuations in $I_{\rm u}$ is incorrect~\cite{mora2008,vitu2008}
 and Eq.~\eqref{current} becomes necessary.

\subsection{Elastic contribution to the current}\label{elastic-current}

We are now in a position to compute the mean value of the current in an 
out-of-equilibrium situation. A dc bias is applied between the two electrodes
imposing $\mu_{\rm L} - \mu_{\rm R} = e V$. Left and right scattering states,
corresponding to $c_{L,k\sigma}$ and $c_{R,k\sigma}$ operators, are in thermal
equilibrium with chemical potentials $\mu_{\rm L}$ and $\mu_{\rm R}$.
Hence, using Eq.~\eqref{rota}, we obtain the populations
\begin{subequations}\label{steady}
\begin{align}
\langle b_k^\dagger b_{k'} \rangle & = \delta_{k,k'} \left[
\cos^2 \theta f_{\rm L} (\varepsilon_k) 
+  \sin^2 \theta f_{\rm R} (\varepsilon_k) \right], \\[1mm]
 \langle a_k^\dagger a_{k'} \rangle &  = \delta_{k,k'} \left[
\sin^2 \theta f_{\rm L} (\varepsilon_k) 
+  \cos^2 \theta f_{\rm R} (\varepsilon_k) \right], \\[1mm]
\langle a_k^\dagger b_{k'} \rangle & = \langle  b_k^\dagger a_{k'} \rangle =
\frac{\delta_{k,k'} \sin  2 \theta }{2}  [ f_{\rm L} (\varepsilon_k) 
- f_{\rm R} (\varepsilon_k) ], \\[1mm]
f_{\rm L/R} (\varepsilon) & = f(\varepsilon - \mu_{\rm L/R}),
\end{align}
\end{subequations}
for all spins $\sigma$. Eq.~\eqref{choche}, that implies a vanishing Hartree diagram,
is satisfied with $\mu_{\rm L} = \sin^2 \theta \, e V$ and $\mu_{\rm R} 
= - \cos^2 \theta \, e V$. $f (\varepsilon) = (1 + e^{\beta \varepsilon})^{-1}$ is
the Fermi distribution.

The average current is obtained from Eq.~\eqref{current} and reproduces the
Landauer-B\"uttiker formula~\cite{blanter2000}
\begin{equation}\label{landauer}
I_{\rm el} = \frac{N e}{h}  \int_{-\infty}^{+\infty} d \varepsilon T(\varepsilon)
[ f_{\rm L} (\varepsilon) - f_{\rm R} (\varepsilon) ],
\end{equation}
with the transmission 
\begin{equation}\label{transmission}
T(\varepsilon) = \sin^2 2 \theta \, \sin^2 (\delta(\varepsilon)),
\end{equation}
and the phase shift
\begin{equation}\label{pshift2}
\delta(\varepsilon) = \delta_0 + \frac{\alpha_1}{T_K} \varepsilon 
+ \frac{\alpha_2}{T_K^2} \left( \varepsilon^2 - \frac{(\pi T)^2}{3} - 
\frac{(e V)^2 \, \sin^2 2 \theta}{4}  \right),
\end{equation}
where we have used the identity~\eqref{FLs2}, $\alpha_2 = (N-1)\, \phi_2 /4$.
Here, the phase shift $\delta(\varepsilon)$ has an extra $(V,T)$ dependence
due to mean-field (Hartree) interaction contributions ({\it cf.} Eq.~\eqref{phase2}).  
Within the heuristic resonant-level picture, we can interpret this as the voltage inducing
a quasiparticle population, whose interactions in turn yield a mean-field upward energy shift of the Kondo resonance.  Note that
the relevant interactions here are not the leading-order Fermi liquid interactions described by $\phi_1$, but rather the next-leading-order
interaction described by $\phi_2$.

At zero temperature, the current can be expanded to second order 
in $e V/T_K$. The asymmetry and the zero-energy transmission are characterized by
\begin{equation}\label{defC}
C = \cos 2 \theta, \qquad T_0 = \sin^2 \delta_0
\end{equation}
with $C=0$ in the symmetric case. The current takes the form
\begin{equation}\label{elacur2}
\begin{split}
& \frac{I_{\rm el}}{(1-C^2) N e^2 V / h} = T_0 - 
C \sin 2 \delta_0 \, \alpha_1 \frac{e V}{2 T_K} \\[2mm]
& +  \left( \frac{e V}{T_K} \right)^2 
\left[ \cos 2 \delta_0 (1+3 C^2) \frac{\alpha_1^2}{12}   
-  \sin 2 \delta_0 (1-3 C^2)  \frac{\alpha_2}{6} 
 \right].
\end{split}
\end{equation}

\subsection{Inelastic contribution to the current}\label{inelastic}

The Keldysh framework~\cite{kamenev2005} is well-suited to estimate 
interaction corrections~\eqref{interaction} to the current. The mean
current takes the form
\begin{equation}\label{current2}
I = \langle T_c \hat{I} (t) e^{-\frac{i}{\hbar} \int_{\cal C} d t' H_{\rm int} (t')}
\rangle,
\end{equation}
where the Keldysh contour ${\cal C}$ runs along the forward 
time direction on the branch $\eta=+$ followed by
a backward evolution on the branch $\eta=-$. $T_c$ is the
corresponding time ordering operator. Time evolution of
$\hat{I} (t)$ and $ H_{\rm int} (t)$ is in the interaction
representation with the unperturbed Hamitonian $H_0$~\eqref{elastic}.
Mean values $\langle \ldots \rangle$ are also taken with
respect to $H_0$~\eqref{elastic} with bias voltage,
see Eqs.~\eqref{steady}.
Note that the time $t$ in Eq.~\eqref{current2} is arbitrary
for our steady-state situation.
Finally, in order to maintain the original order of operators
in $\hat{I} (t)$, we take left (creation) operators on 
the $\eta=-$ branch and right (annihilation) one 
on the $\eta=+$ branch.

\begin{figure}
\includegraphics[width=4.cm]{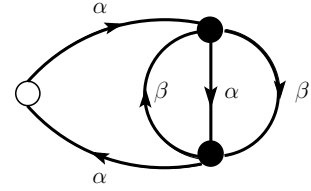}
\caption{\label{self-three} Second order diagram describing the interaction
correction to the current from  Hamiltonian Eq.~\eqref{interaction}.
The open circle represents a
current vertex while filled black dots correspond to interaction
vertices. $\alpha,\beta$ are spin degrees of freedom.
The self-energy term is formed by the three lines connecting the two
interaction vertices.}
\end{figure}

A perturbative study of Eq.~\eqref{current2} is possible
by expansion in $H_{\rm int}$ and use of Wick's theorem.
This leads to usual diagrammatics where one should keep
track of the Keldysh branch index.
The lowest order recovers the results of 
Sec.~\ref{elastic-current} describing elastic scattering. 
The next first order 
gives only Hartree terms already included in Eq.~\eqref{elacur2}.
$H_{\rm int}$ gives rise in general to three vertices with 
coefficients $\phi_1$, $\phi_2$ and $\chi_2$ where the last two
are already second order in $1/T_K$. Thus it is consistent to
keep only $\phi_1$ in the second order expansion in $H_{\rm int}$. 
A typical Green's function is defined by
${\cal G}_{ab}^{\eta_1,\eta_2} (x-x',t-t') = - i \langle T_c 
a (x,t,\eta_1) b^\dagger (x',t',\eta_2) \rangle$.   
For clarity, spin indices are omitted here and below  since all noninteracting
Green's functions are spin diagonal.
Noninteracting Green's functions are $2 \times 2$ matrices in Keldysh
space  given in momentum-energy space by
\begin{subequations}\label{nong}
\begin{align}
\label{Gbb}
{\cal G}_{bb} (k,\varepsilon)
& =  \frac{1}{\varepsilon-\varepsilon_k} \tau_z + i \pi \begin{pmatrix} F_0 & F_0+1
\\ F_0  - 1 & F_0
\end{pmatrix} \delta(\varepsilon - \varepsilon_k) \\[2mm]
{\cal G}_{aa} (k,\varepsilon)
& =  \frac{1}{\varepsilon-\varepsilon_k} \tau_z + i \pi \begin{pmatrix} \tilde{F}_0 
& \tilde{F}_0+1
\\ \tilde{F}_0  - 1 & \tilde{F}_0 
\end{pmatrix} \delta(\varepsilon - \varepsilon_k) \\[2mm]
{\cal G}_{ab}  (k,\varepsilon) & = {\cal G}_{ba} (k,\varepsilon) 
= 2 i \pi \langle a^\dagger_k b_k \rangle  \begin{pmatrix}
1 & 1 \\ 1 & 1  \end{pmatrix} \delta(\varepsilon - \varepsilon_k),
\end{align}
\end{subequations}
with the Pauli matrix $\tau_z = \begin{pmatrix}
1 & 0 \\ 0 & -1  \end{pmatrix}$, $F_0 (\varepsilon_k) = 2 \langle b_k^\dagger b_k \rangle - 1 $,
and $\tilde{F}_0 (\varepsilon_k) = 2 \langle a_k^\dagger a_k \rangle - 1 $,
as given Eqs.~\eqref{steady}.
We wish to compute the second order correction from Eq.~\eqref{current2}.
It involves the self-energy contribution 
shown Fig.\ref{self-three} and defined by
\begin{equation}\label{sigma}
\begin{split}
\Sigma^{\eta_1,\eta_2}&  (t_1-t_2)  = \sum_{k_1,k_2,k_3} 
{\cal G}^{\eta_1,\eta_2}_{bb} (k_1,t_1-t_2) \times \\[1mm]
& \times {\cal G}^{\eta_2,\eta_1}_{bb} (k_2,t_2-t_1) 
{\cal G}^{\eta_1,\eta_2}_{bb} (k_3,t_1-t_2)
\end{split}
\end{equation}
The {\it causality} identity, for $t \ne 0$,
\begin{equation}\label{causal}
\Sigma^{++} (t) +  \Sigma^{--} (t) = \Sigma^{+-} (t) +  \Sigma^{-+} (t),
\end{equation}
is derived by writing the explicit time dependence in Eq.~\eqref{sigma}.
It leads to various cancellations, in particular for terms where
the lines external to the self-energy~\eqref{sigma} bear no $\eta_{1/2}$
dependence. The lines that join the current vertex to the self-energy
in Fig.\ref{self-three} travels from $x$ (or $-x$) to $0$ (the dot) and the opposite.
Thus, using the Green's function~\eqref{Gbb} in real space (with $\alpha = \pm 1$)
\begin{equation}
{\cal G}_{bb}^{\eta_1,\eta_2} (\alpha x,\varepsilon) = i \pi \nu e^{i \alpha \varepsilon x/v_F} 
\left( F_0(\varepsilon) + \begin{cases}
+ \eta_1 \quad \alpha = 1 \\
- \eta_2 \quad \alpha = -1 
\end{cases} \! \! \! \! \! \! \right)
\end{equation}
and the identity~\eqref{causal}, one shows that the terms with operators
taken at $x$ in Eq.~\eqref{current} give a vanishing contribution
to the current. This is merely a consequence of causality: interaction,
which takes place at $x=0$, can only affect outgoing current and not
the incoming part.
We are left with the current correction
\begin{equation}\label{curr1}
\begin{split}
& \delta I_{\rm int}  = 
\frac{N (N-1) e \sin 2 \theta}{2 \nu h}  
\left( \frac{\phi_1}{\pi \nu^2 T_K} \right)^2 \sum_{\eta_1,\eta_2} \eta_1
\eta_2 \times \\[2mm]
&  \int \frac{d \varepsilon}{2 \pi} 
( i {\cal S} {\cal G}_{bb}^{+,\eta_1}  (-x,\varepsilon) \Sigma^{\eta_1,\eta_2} (\varepsilon)
{\cal G}_{ba}^{\eta_2,-} (x,\varepsilon) + c.c. ).
\end{split}
\end{equation}
The summation over $\eta_1$ and $\eta_2$ gives two terms: (i)
one includes the combination $\Sigma^{++} - \Sigma^{--}$. It gives
a contribution proportional to $D$ exactly cancelled by a counterterm.
Details are given in Appendix~\ref{counterterms}. 
(ii) the second term involves
the combination $\Sigma^{+-} - \Sigma^{-+}$ and remains finite in
the limit  $D \to +\infty$.
It reads
\begin{equation}\label{int-inel}
\begin{split}
& \delta I_{\rm int}  = 
 \frac{N (N-1) (1-C^2) e \pi}{2\, h}  
\left( \frac{\phi_1}{\pi \nu^2 T_K} \right)^2 
\times \\[2mm]
& ( {\cal S} + {\cal S}^*)  \int \frac{d \varepsilon}{2 \pi} 
(\Sigma^{-+} - \Sigma^{+-}) (\varepsilon) \, i \pi \nu \Delta f (\varepsilon),
\end{split}
\end{equation}
with $C$ given by Eq.~\eqref{defC} and $\Delta f (\varepsilon) = 
f_{\rm L} (\varepsilon) - f_{\rm R} (\varepsilon) $.

We proceed further and restrict ourselves to the zero-temperature case.
The left and right Fermi step functions are introduced by
going to frequency space for Eq.~\eqref{sigma}, and then
by using Eqs.~\eqref{Gbb} and~\eqref{steady}. 
The result involves a sum of terms with products of $\cos^2 \theta$
and $\sin^2 \theta$. Two distinct integrals,
\begin{subequations}\label{Js}
\begin{align}
J_1 &=  \int_{\mu_{\rm L}} d \varepsilon
\int_{\mu_{\rm R}} d \varepsilon' \int^{\mu_{\rm L}} d \varepsilon^{\prime \prime}
f_{\rm L} ( \varepsilon + \varepsilon' - \varepsilon^{\prime \prime}) 
 \\[1mm]
J_2 &= \int_{\mu_{\rm R}} d \varepsilon
\int_{\mu_{\rm R}} d \varepsilon' \int^{\mu_{\rm L}} d \varepsilon^{\prime \prime}
f_{\rm L} ( \varepsilon + \varepsilon' - \varepsilon^{\prime \prime})
\end{align}
\end{subequations}
corresponding respectively to one- and two-particles transfer,
appear with the following combination
\begin{equation}
\cos^2 \theta \sin^2 \theta \, ( J_2 -2 J_1) + J_1
= \frac{J_2 (1-C^2) + 2 J_1 (1+C^2)}{4}.
\end{equation}
With $J_1 = (e V)^3/6$ and $J_2 = 4 (e V)^3/3$, we obtain the
current correction
\begin{equation}\label{intercorr}
\frac{\delta I_{\rm int}}{(1-C^2) N e^2 V / h}
= \cos 2 \delta_0 \,(N-1) \left( \frac{\phi_1 \, e V}{ T_K} \right)^2
\left( \frac{5}{12} - \frac{C^2}{4} \right).
\end{equation}

This result can be given a quite simple physical interpretation
along the line of Ref.~\cite{sela2006}. The $\phi_1$ term in the 
interaction part
of the Hamiltonian~\eqref{interaction} can be decomposed 
on the left/right operators basis using Eq.~\eqref{rota}.
It then describes processes where $0$, $1$ or $2$ electrons
are transfered from one scattering state to the other.
Using Fermi's golden rule and $\cos^6 \theta \sin^2 \theta
+ \cos^2 \theta \sin^6 \theta = (1-C^4)/8$,
the total rate of one-electron transfer is evaluated to be 
$2 \Gamma_1 (1-C^4)$ where 
\begin{equation}
\Gamma_1 = N (N-1) \frac{e V}{h} \frac{\phi_1^2}{24} 
\left(\frac{e V}{T_K} \right)^2.
\end{equation}
From $\cos^4 \theta \sin^4 \theta = (1-C^2)^2/16$, the total rate
for two-electron transfer is $\Gamma_2 (1-C^2)^2/2$ 
where $\Gamma_2 = 8\Gamma_1$. For one- and two-electron transfers,
$e \cos 2 \delta_0$ and $2 e \cos \delta_0$
are interpreted as the corresponding charge transfered between leads~\cite{vitu2008}.
Writing the current correction as 
\[
\delta I_{\rm int} = 
(e \cos 2 \delta_0) 2 \Gamma_1 (1-C^4) 
+ (2 e \cos \delta_0) \frac{\Gamma_2}{2} (1-C^2)^2,
\]
we recover Eq.~\eqref{intercorr}.

\subsection{Current for SU(2) and SU(4)}

The results of Secs.~\ref{elastic-current},~\ref{inelastic}
can be extended to finite temperature as explained in Appendix~\ref{finiteT}.
We detail results for the total current $I = I_{\rm el}
+ \delta I_{\rm int}$ in the $(N=2, m=1)$ case and 
$(N=4, m=1,2)$ cases. 

For SU(2), a single electron is trapped on the dot, 
$\alpha_1 = \phi_1$ and $\alpha_2=0$. The current takes
the form
\begin{equation}\label{cur1}
I =  I_{\rm m} \left[ 1 - \left(\frac{\alpha_1}{T_K}\right)^2
\left( \frac{(e V)^2}{2} + (\pi T)^2 \right) \right],
\end{equation}
where $I_{\rm m} = (2 e^2 V/h) (1-C^2)$.
In the particle-hole SU(4) symmetric case with two electrons,
$\alpha_1 = 3 \phi_1$ and $\alpha_2 = 0$.
The current reads
\begin{equation}\label{cur2}
I =  I_{\rm m} \left[ 1 - \left(\frac{\alpha_1}{T_K}\right)^2
\left( \frac{2 (e V)^2}{9} + \frac{C^2 (e V)^2}{6}
 + \frac{5 (\pi T)^2}{9} \right) \right],
\end{equation}
where $I_{\rm m} = (4 e^2 V/h) (1-C^2)$.

Turning now to the SU($4$) case with one electron on the dot, one 
finds that the inelastic contribution to the current vanishes identically (c.f. Eq.~\eqref{intercorr}), as the `effective charges' associated with
interaction-induced scattering events are proportional to $\cos 2 \delta_0$ and hence identically zero \cite{vitu2008}.  
The only contribution is thus from the elastic channel (c.f. Eq.~\eqref{elacur2}), yielding:
\begin{equation}\label{isu4}
I =  I_{\rm m}
 \left[ 1 - \frac{\alpha_1 C e V}{T_K} - 
\frac{\alpha_2}{3} \left(\frac{e V}{T_K}\right)^2
(1 - 3C^2) \right],
\end{equation}
where $I_{\rm m} = (2 e^2 V/h) (1-C^2)$. There is no temperature correction up to this order
of the low energy expansion. The case with three electrons $(m=3)$ and SU(4) symmetry
 is related to the one-electron
case by particle-hole symmetry. The Kondo resonance is thus changed from above
to below the Fermi energy.
The result for the current is then the same as Eq.~\eqref{isu4},
but with an opposite sign for the asymmetry  ($\theta \to \pi/2-\theta$, $C \to -C$), {\it i.e.}
the roles of left (L) and right (R) leads are exchanged for hole transport. 

The differential conductance $G (V) = \frac{d I}{d V} (V)$ 
obtained from Eq.~\eqref{isu4} gives an asymmetric curve
whenever $C\ne 0$. 
Consider the first the strongly asymmetric case, where $|C|$ becomes  sizeable.  In this case,
the asymmetric linear $e V/T_K$ correction in Eq.~\eqref{isu4}
dominates even at low bias voltage. For strong asymmetry $|C| \to 1$,
the conductance measures the density of states of the Kondo resonance~\cite{glazman2005}
at $\pm e V$.  The asymmetric linear term thus follows the side of the Kondo resonance
and reveals that the resonance peak is located away from the Fermi level~\cite{lehur2007}.
This behaviour is in fact generic to the SU(N) case when the occupation of the dot
is away from half-filling. In the SU(2) case or generally for a half-filled dot ($m=N/2$),
the resonance peak is located at the Fermi level which suppresses the asymmetric
linear term, see Eqs.~\eqref{cur1} and~\eqref{cur2}.

Turning now to the case of a symmetric dot-lead coupling ($C=0$), we see that as expected, the 
differential conductance $G (V)$ is symmetric in $V$ at all dot 
fillings; hence, it exhibits a quadratic behaviour at low bias.  In the SU($4$) case, the 
conductance obtained from Eq.~\eqref{isu4} is predicted to be maximum at $V=0$, in agreement with
results obtained from slave boson mean field theory~\cite{delattre2009}.  Within the Fermi liquid approach, and for one electron on the dot, this behaviour
is at first glance rather puzzling.  As we have already indicated, in the SU($4$) case, the conductance is completely due to the
elastic transport channel.  Using the heuristic picture provide by the resonant level picture (i.e.~elastic scattering due to a Lorentzian Kondo resonance sitting above the Fermi energy), one would expect that the differential conductance should {\it increase} with increasing voltage, due to the positive curvature of the expected (Lorentzian) transmission coefficient.  This picture is in fact incorrect, as it neglects the important Hartree contributions discussed in Sec.~\ref{elastic-current}.  Heuristically, as the voltage is increased, quasiparticle interactions lead to a mean-field upward energy shift of the position of the Kondo resonance.  Because of the relation $\phi_2 = (4/3) \alpha_2$, this energy-shift effect dominates, and causes the conductance to decrease; without this mean-field energy shift, the conductance would indeed exhibit a quadratic increase at small voltages.  Note that an incorrect upturn in the conductance was reported in previous works:  Ref. \cite{vitu2008} neglected the higher-order Fermi liquid interaction parameter $\phi_2$ and the resulting mean-field energy shift, while Ref.~\cite{mora2008} treated it incorrectly (corrected in \cite{mora2009}).  Note also that the results for the conductance presented in Ref.~\cite{lehur2007} only apply to a system with a strongly asymmetric dot-lead coupling.

\section{Current noise}\label{currentnoise}

Fluctuations in the current are almost as important as the current itself.
In particular, the shot noise (at zero temperature) carries information about
charge transfer in the mesoscopic system.
The purpose of this section is to detail the calculation of the zero-frequency
current noise,
\begin{equation}\label{noise}
S \equiv 2 \int dt \langle \Delta \hat{I} (t) \Delta \hat{I}(0) \rangle,
\end{equation}
with the current fluctuation $\Delta \hat{I} (t) = \hat{I} (t) 
- \langle \hat{I} (t)  \rangle$, see Eq.~\eqref{current} for the current 
operator expression.

Insight can be gained by first examining the strong coupling fixed 
point at zero temperature, with $e V \ll T_K$ so that $\delta (\varepsilon) \simeq \delta_0$. 
Quantum expectations in Eq.~\eqref{noise} are evaluated 
with the free Hamiltonian~\eqref{elastic}. The shot noise,
\begin{equation}\label{linear}
S_0 = \frac{2 N e^3 |V|}{h}  T_0 (1-C^2) [  1- T_0 (1-C^2) ],
\end{equation}
is  pure partition noise like a coherent scatterer~\cite{blanter2000}.
This result implies a vanishing noise in the particle-hole symmetric case,
like standard SU(2), with symmetric leads coupling ($T_0=1$ and $C=0$).
In this specific case, the shot noise is only determined by the 
vicinity of the Kondo strong coupling fixed point, that is by the inelastic
Hamiltonian~\eqref{interaction} and the corrections to $\delta_0$
in the elastic phase shift~\eqref{pshift}. The shot noise is therefore
highly non-linear with $S \sim V^3$ at low bias voltage.
Since the corresponding current is close to unitarity, an effective charge
$e^* = (5/3) \, e$ has been extracted from the ratio of the noise 
to the backscattering  current~\cite{sela2006}. $e^* \ne e$ should however 
not be confused with a fractional charge. It emerges as an average charge
during additional and independent Poissonian processes involving one and two charges 
transfer as shown by
the calculation of the full couting statistics~\cite{gogolin2006b}.
Nevertheless, this charge $e^* = (5/3) \, e$ is universal and characterizes
the vicinity of the Kondo strong coupling fixed point. It can be seen as an 
out-of-equilibrium equivalent of the Wilson ratio.

In asymmetric situations ($T_0 \ne 1$ or $C \ne 0$), the linear 
part~\eqref{linear} of the noise does not vanish and even dominates at low
bias voltage. For instance 
in the SU(4) case, $T_0 = 1/2$ so that $T_0 (1-T_0) = 1/4$.
This property is quite relevant for experiments and may be used to 
discriminate SU(2) and SU(4) symmetries for which the current gives
essentially the same answer~\cite{delattre2009}.
In a way similar to the symmetric SU(2) case, we can define an effective
charge from the ratio of the non-linear parts ($\sim V^3$) in the noise
and the current~\cite{vitu2008,mora2008}. 
This is however less straightforward to measure
experimentally since it requires a proper subtraction of the linear terms.

\subsection{Elastic contribution to the noise}\label{elastic-noise}

Inserting the current operator~\eqref{current} in Eq.~\eqref{noise},
the elastic Hamiltonian~\eqref{elastic} gives a gaussian measure
which allows to use Wick's theorem, and thus Eqs.~\eqref{steady}.
Like for the current, we obtain a Landauer-B\"uttiker formula~\cite{blanter2000}
for the noise with the same transmission~\eqref{transmission}
and phase shift~\eqref{pshift2}. At zero temperature, it reads
\begin{equation}
S = \frac{2 N e^2}{h} \int_{\mu_{\rm R}}^{\mu_{\rm L}} d \varepsilon \, T (\varepsilon)
[ 1 -  T (\varepsilon) ].
\end{equation}
An expansion to second order in  $e V/T_K$ yields the elastic (non-linear) 
correction to the noise~\eqref{linear},
\begin{equation}\label{dsel}
\frac{\delta S_{\rm el}}{ (1-C^2) 2 N e^3 | V | /h} =  \delta \bar{s}_{\rm el}^{(1)}
\frac{e V}{T_K}
+ \delta \bar{s}_{\rm el}^{(2)} \left(  \frac{e V}{T_K} \right)^2  ,
\end{equation}
with coefficients,
\begin{subequations}
\begin{align}
\label{first-o}
\delta \bar{s}_{\rm el}^{(1)}
& = - \frac{C \alpha_1 \sin 2 \delta_0}{2}  \, [ 1 - 2 T_0 (1-C^2) ], \\[1mm]
\begin{split}
\delta \bar{s}_{\rm el}^{(2)} & = \frac{\alpha_1^2}{12} (1+3 C^2) ( \cos 4 \delta_0 +
2 \sin \delta_0 \sin 3 \delta_0 C^2 ) \\[1mm]
& - \frac{\alpha_2}{6} (1 - 3 C^2) \sin 2 \delta_0 \, [ 1 -  2 T_0 (1 - C^2) ],
\end{split}
\end{align}
\end{subequations}
and the total elastic noise reads $S_{\rm el} = S_0 + \delta S_{\rm el}$.
The first order correction~\eqref{first-o} gives an asymmetric 
part to the noise for $C \ne 0$.
In a way similar to the current case, 
particle-hole transformation ($\delta_0 \to \pi - \delta_0$, $\alpha_2 \to -\alpha_2$)
reverts the sign of the asymmetry~\eqref{first-o} 
which indicates that the Kondo resonance is centered off the Fermi level.

\subsection{Inelastic contribution to the noise}

\begin{figure}
\includegraphics[width=5.5cm]{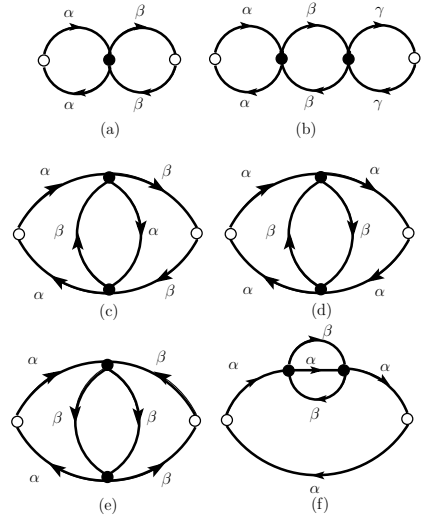}
\caption{\label{diag} Diagrams for the noise appearing in
the first and  second order expansions in the inelastic Hamiltonian Eq.~\eqref{interaction}.
Diagrams (c) and (d) give identical contributions.
For diagram (f), the three-lines bubble can alternatively dress the
bottom Green's function.  $\alpha,\beta,\gamma$ denote spins with
$\alpha \ne \beta$ and $\beta \ne \gamma$. Open dots represent 
current vertices while filled black dots correspond to interaction
vertices.}
\end{figure}

We follow the same procedure as for the interaction correction to the current 
established in Sec.~\ref{inelastic}. The mean value in Eq.~\eqref{noise} is taken
within the Keldysh framework, similar to Eq.~\eqref{current2}. The correct ordering
of $\hat{I}$ operators is maintained by choosing time $0$ on the $\eta = +$ branch
and time $t$ on the $\eta = -$ branch. The perturbative study of the noise
involves diagrams with two current vertices instead of one in Sec.\ref{inelastic}.
The resulting calculations are therefore similar to those for the current but
are much more involved on the technical side.
The diagrams relevant for the noise at first and second order in $1/T_K$ are shown
Fig.\ref{diag}.
Noninteracting Green's functions are still given by Eqs.~\eqref{nong}.

Three vertices can be built from the interaction 
Hamiltonian $H_{\rm int}$~\eqref{interaction}  with coefficients 
 $\phi_1$, $\phi_2$ and $\chi_2$. The $\chi_2$ vertex has six legs
and appears at most once at order $1/T_K^2$. Topology therefore imposes
that two legs among the six must connect to form a closed loop.
The corresponding energy integral vanishes thanks to Eq.~\eqref{choche}.
Apart from Hartree terms already included in the elastic 
part, see Sec.\ref{elastic-noise}, the expansion to first order
in $H_{\rm int}$ gives the single diagram~\ref{diag}.a. 
Involving both $\phi_1$ and $\phi_2$, the interaction vertex in 
diagram~\ref{diag}.a is characterized by
the energy dependent coefficient $\phi(\varepsilon,\varepsilon')
= \phi_1 + \phi_2 (\varepsilon + \varepsilon')/2 T_K$.
The corresponding noise correction is given by
\begin{equation}
\begin{split}
\delta S_{\rm int}^a &= \frac{-i e^2 (1-C^2) N (N-1)}{4 \pi^2 \nu^4 h T_K} \\[1mm]
& \times \sum_{\eta} \eta \int \frac{d \varepsilon d\varepsilon'}{(2 \pi)^2} 
A^{\eta\eta}(\varepsilon) \phi(\varepsilon,\varepsilon')  A^{\eta\eta}
(\varepsilon'),
\end{split}
\end{equation}
where 
 $A^{\eta_1,\eta_2}$ is the building block
defined in Appendix~\ref{details} and shown in Fig.\ref{building}.
We replace  $A^{\eta_1,\eta_2}$ by its expression~\eqref{defA} (with $\sin^2 (\delta(\varepsilon))$
instead of $T_0$) and perform the $\eta$ summation to get
\begin{equation}
\begin{split}
& \frac{\delta S_{\rm int}^a}{(1-C^2)^2 2 N e^2/h}  =  \frac{(N-1)}{T_K} 
 \int d \varepsilon d \varepsilon' \Delta f(\varepsilon) \\[1mm] & \times 
 F_0 (\varepsilon) \sin^2 (\delta(\varepsilon)) \phi(\varepsilon,\varepsilon')
\Delta f(\varepsilon') \sin 2 \delta(\varepsilon').
\end{split}
\end{equation}
This general expression can finally be expanded order by order in $e V/T_K$.
After energy integration, we obtain at zero temperature a first 
and a second order noise term, 
\begin{equation}\label{dsa}
\frac{\delta S_{\rm int}^a}{ (1-C^2)^2 2 N e^3 | V | /h} =  \delta \bar{s}_{\rm int}^{(a,1)}
\frac{e V}{T_K}
+ \delta \bar{s}_{\rm int}^{(a,2)} \left(  \frac{e V}{T_K} \right)^2  ,
\end{equation}
with coefficients,
\begin{subequations}
\begin{align}
\delta \bar{s}_{\rm int}^{(a,1)}
& = C \phi_1 (N-1)  \sin 2 \delta_0 \, \sin^2 \delta_0, \\[1mm]
\begin{split}
\delta \bar{s}_{\rm int}^{(a,2)} & = - C^2 (N-1) [
\phi_2 T_0 \sin 2 \delta_0 \\[1mm]
& + 6 \alpha_1 \phi_1 T_0 (1-4 T_0/3) ]/2.
\end{split}
\end{align}
\end{subequations}
Note that these two terms vanish identically for symmetric leads coupling
($C=0$). It can be checked again that the first order correction is  odd
with respect to particle-hole symmetry while the second order is even.

The expansion to second order in $H_{\rm int}$ yields the diagrams~\ref{diag}.b-f
with two interaction vertices. To be consistent with the rest of the perturbative
calculation, only $\phi_1$ is kept in each interaction vertex.
The contributions corresponding to the diagrams~\ref{diag}.b-f are all calculated
in Appendix~\ref{details}. Finally, the total noise reads
\begin{equation}
S = S_0 + \delta S_{\rm el} + \delta S_{\rm int}^a + \delta S_{\rm int}^b
+ 2 \delta S_{\rm int}^c + \delta S_{\rm int}^e + \delta S_{\rm int}^f,
\end{equation}
where the different terms are respectively given by Eqs.~\eqref{linear},
\eqref{dsel}, \eqref{dsa}, \eqref{dsb}, \eqref{dsc}, \eqref{dse} and~\eqref{dsf}.

\subsection{Noise for SU(2) and SU(4)}

We have also extended the noise calculation to finite temperature along the lines
of Appendix~\ref{finiteT}. In the asymmetric case, the results are too cumbersome
to be written here. In the symmetric case, the noise was calculated in 
Ref.\cite{mora2008} where it was emphasized that corrections are 
rapidly sizeable at finite temperature. Hence the shot noise
regime is expected only at very low temperature.
Keeping a zero temperature, we specialize here to the experimentally relevant
SU(2) and SU(4) cases with one electron on the dot, $m=1$.

In the SU(2) case, the noise correction to Eq.~\eqref{linear} reads
\begin{equation}\label{noisecorr}
\frac{\delta S}{(1-C^2) 2 N e^3 |V| /h} = \left( \frac{e V}{T_K} \right)^2
\alpha_1^2 \left(\frac{5}{6} - \frac{4}{3} \, C^2 \right),
\end{equation}
where the $C^4$ terms cancel each other unexpectedly.

In the SU(4) case, the noise correction has linear and quadratic contributions,
\begin{equation}
\begin{split}
& \frac{\delta S}{(1-C^2) 2 N e^3 |V| /h} = \left( \frac{e V}{T_K} \right)
\frac{C \alpha_1}{2} (1- 2 C^2)  \\[1mm]
& + \left( \frac{e V}{T_K} \right)^2
\left[ \frac{\alpha_1^2}{18} (1 - 8 C^2 + 7 C^4) - \frac{\alpha_2}{6} (1 - 3 C^2)
\right].
\end{split}
\end{equation}

\section{Results and conclusion}
\label{results}

\subsection{Main Results}

Following Refs~\cite{vitu2008,mora2008}, we define a generalized Fano factor
$F$ which describes the relation between the nonlinear current and current noise:
\begin{equation}\label{fano}
F = \frac{1}{2 e} \frac{\delta S}{\delta I}.
\end{equation}
It is defined as the ratio between the non-linear parts of the noise
$\delta S = S - S_0$ (c.f.~Eq.~\eqref{linear})
and of the current $\delta I  =  I - I_0$, where
\begin{equation}
I_0 = (1 - C^2) \frac{N T_0 e^2 V}{h},
\end{equation}
is the linear current (for $e V \ll T_K$).
We focus on the non-linear noise and current, as it is these quantities which are sensitive to
the contribution of Fermi liquid interactions.  

Consider first the strong asymmetric case 
$|C| \to 1$ ({\it i.e.}~$\theta \to 0$ or $\theta \to \pi/2$), where 
the dot is strongly coupled to one lead and only weakly to the other (c.f.~
Eq.~\eqref{rota}).  Transport in this limit corresponds to 
an incoherent tunneling regime where the hopping
from the weakly-coupled lead to the dot is the  limiting process.
It can be checked from the Eqs.~\eqref{dsel}, \eqref{dsa}, 
\eqref{dsb}, \eqref{dsc}, \eqref{dse}, \eqref{dsf} for the noise, and
Eqs.~\eqref{elacur2}, \eqref{intercorr} for the current, that the Fano factor $F=1$ to
leading order in $1-|C|$. This is of course expected since the tunneling
regime gives  Poissonian statistics for charge transfer.
Note that this unity ratio holds order by order for the $e V/T_K$ and $(eV/T_K)^2$ correction separately.
In addition, we also have
$S_0/ 2 e I_0 = 1$ to leading order in $1-|C|$.

In the opposite limit of a symmetric dot-lead couping (i.e. $C=0$), 
coherent effects are important to transport, and charge transport is generally
not Poissonian~\cite{gogolin2006b}.
Note also that in the symmetric case, the non-linear parts of both 
the current and current noise are $\propto V^3$.  We find that
the generalized Fano factor~\eqref{fano}
 is given by
\begin{equation}\label{fanogen}
F = \frac{1 + \sin^2 (2 \delta_0) + \frac{9-13 \sin^2 (2 \delta_0)}{N-1}
- \frac{\alpha_2}{\alpha_1^2} \sin 4 \delta_0}
{\frac{N+4}{N-1} \, \cos 2 \delta_0 - 2  \frac{\alpha_2}{\alpha_1^2}
\sin 2 \delta_0}.
\end{equation}
This Fano factor includes the effect of interactions; we have used the important equalities in Eqs. (\ref{FLs}).  
Note that this result has no explicit dependence on $V/T_K$:  it is thus
a universal quantity characterizing the Fermi liquid properties of the strong-coupling fixed point; also note that  $F$ is invariant under a particle-hole transformation, where
$m \to N-m$.  We stress that the fact $F \neq 1$ in general is due {\it both} to the existence of two-particle scattering at the fixed point, as well as to the partition noise associated with single-particle scattering.  We give in  Table~\ref{table} values of $F$ for different $N$ and $m$.

\begin{table*}
\begin{tabular}{|c||*{8}{r|}}
\hline & \multicolumn{8}{c|}{N } \\
\cline{2-9}
m & 2 & 3 & 4 &5 &6 &7 &8 &9 \\
\hline
1 & -5/3 & -0.672 & -0.300 & -0.156 & 0.003 & 0.156 & 0.287 & 0.393 \\
\hline
2 &  &  -0.672 & -3/2 & -1.256 & -1.031 & -0.855 & -0.679 & -0.503 \\
\hline
3 &  &  & -0.300 & -1.256 & -7/5 & -1.326 & -1.254 & -1.173 \\
\hline
4 &  &  &  & -0.156 & -1.031 & -1.326 & -4/3 & -1.313 \\
\hline
\end{tabular}
\caption{Fano factor $F$, Eq.~\ref{fano}, for various $N$ and $m$.}
\label{table}
\end{table*}

For $N \to +\infty$, Eq.~\eqref{fanogen} leads to 
\begin{equation}\label{fanoN}
F = \frac{3 \cos 4 \delta_0 + 4 \cos 2 \delta_0 - 1}{4 + 2 \cos 2 \delta_0}.
\end{equation}
Note that in the large $N$ limit, two-particle scattering processes become insignificant~\cite{ratiani2009} for the current (since
$\phi_1$ and $\phi_2$ scale as $\sim 1/N$) and the result is consistent with the non-interacting resonant level. 
In this limit, the Wilson ratio is in fact just one~\cite{nozieres1980,newns1987}.
However, the effect of two-particle scattering processes seems to survive in the current noise through the diagram of Fig.~\ref{diag}(b).
Heuristically, this diagram represents an enhancement of the coherent partition noise already present in the absence 
of Fermi liquid interactions.
The small interaction parameter $\phi_1^2 \sim 1/N^2$ is compensated by the spin summation with $\sim N^3$ equivalent
diagrams. The effect is therefore linear in $N$, at the same level as elastic terms.

The expression of Eq.~\eqref{fanoN} can be checked in two limiting cases.
For $\delta_0 \to 0$, it gives $F\simeq 1$. Again it corresponds to the
tunneling regime since a 
 small phase shift $\delta_0$ implies
a weak electronic transmission $T_0 = \sin^2 \delta_0$. 
When particle-hole symmetry is recovered, $\delta_0 = \pi/2$, we find $F = -1$.
In this limit, the conductance is close to unitarity and interactions play no role since
the diagram of Fig.~\ref{diag}(b) gives a vanishing contribution for $\delta_0 = \pi/2$.
The situation is therefore similar to the ordinary SU(2) 
case~\cite{sela2006,gogolin2006b} where one has Poissonian weak backscattering events.
In our case though, backscattering events are elastic and imply the transfer
of only one electron so that  $F=-1$.

We finally turn to the general asymmetric case, $C \ne 0$, where we focus
on the SU(2) and SU(4) symmetries with $m=1$.
For SU(2), the generalized Fano factor
\begin{equation}\label{fano1}
F = - \frac{5}{3} + \frac{8}{3}\, C^2,
\end{equation}
is obtained from the ratio of the noise~\eqref{noisecorr} and current~\eqref{cur1}
 corrections at zero temperature. We stress that this simple result~\eqref{fano1}
is exact and is not restricted to small values of the asymmetry $C$.
Eq.~\eqref{fano1} indeed bridges the symmetric result $F=-5/3$~\cite{sela2006} to the
tunneling regime, $F=1$ in the strong asymmetry limit $C \to 1$. 
A different asymmetry correction was
predicted in Ref.~\cite{sela2006}. This discrepancy may come from the fact
that the current expression used in Ref.~\cite{sela2006} is not valid outside
the symmetric case $C=0$ (see discussion at the end of Sec.~\ref{subsec-current}).

The SU(4) case for arbitrary asymmetry is more complicated  since the generalized Fano factor~\eqref{fano} bears a $e V/T_K$ dependence.  This is because
the non-linear current and  noise have both linear and quadratic corrections in $e V/T_K$ 
(resp. quadratic and cubic terms in $V$)
and no simplification occurs when the ratio is computed  (universality is however recovered 
in the symmetric case where the linear corrections vanish).
We therefore prefer to compute directly the ratio of the quadratic corrections
with the result
\begin{equation}\label{fanosecond}
F^{(2)} = \frac{1}{2 e} \frac{\delta S^{(2)}}{\delta I^{(2)}}
= - \frac{\alpha_1^2}{3 \alpha_2} \, \frac{1- 8 C^2 + 7 C^4}{1- 3 C^2} + C^2,
\end{equation}
where $\delta S^{(2)}$ ($\delta I^{(2)}$) denotes the noise (current) correction
to second order in $e V/T_K$. Again the ratio~\eqref{fanosecond} connects
the symmetric case ($N=4$, $m=1$ in Table~\ref{table}), $F^{(2)} = - \frac{\alpha_1^2}{3 \alpha_2} \simeq -0.300$~\cite{mora2009}
to the tunnel or strongly asymmetric regime where $F^{(2)}=1$.
Expanding Eq.~\eqref{fanosecond} in $C$, we obtain
$F^{(2)} \simeq -0.300 \, (1 - 8.33 C^2)$ which indicates an important 
correction due to the asymmetry of the coupling to the leads.

\subsection{Conclusion}

To summarize, we have provided a thorough analysis of the non-equilibrium transport in the SU($N$) Kondo regime using an elaborate Fermi-liquid approach. 

We have particularly focused on the case $N=4$ relevant to carbon nanotube quantum dots. One important characteristics of the emergent SU($4$) symmetry is the sign change of the leading current corrections ({\it i.e.} linear in $e V/T_K$) 
as a function of the bias voltage when progressively tuning the asymmetry between the dot-lead couplings. More precisely, for a strong asymmetry, we have recovered a positive linear correction
which traduces the fact that the Kondo resonance is peaked away from the Fermi level; in this case, the conductance measures the density of states of the Kondo resonance at $\pm e V$ where the sign changes with the weakly coupled lead. For symmetric couplings, we have demonstrated that the linear correction now becomes exactly zero and that the current becomes maximum at $V=0$ due to  interactions via the Hartree contributions. In addition, the noise exhibits a non-trivial form due to
the interplay between coherent shot-noise and noise arising from interaction-induced scattering events.
In the symmetric case, interactions result in a universal Fano factor $F\approx -0.300$ at zero temperature. For a finite asymmetry between dot-lead couplings, the current
and the noise have both linear and quadratic corrections in $eV/T_K$. Focusing exclusively on the quadratic corrections, 
we have derived a formula for the Fano factor which extrapolates between the symmetric result
and the strongly asymmetric result $F=1$, perfectly reproducing the Poissonian statistics for charge transfer in the tunneling limit.

In the context of the standard SU($2$) Kondo effect, we have obtained a generalized Fano factor
$F=-5/3+8C^2/3$ at zero temperature which is not restricted to small values of the asymmetry. Finally, in the limit of large $N$, it is certainly relevant to observe that the effect of interactions tends to subsist in the current noise.

{\bf Acknowledgments:} The authors are grateful to M.-S. Choi,  T. Kontos, 
 and N. Regnault for interesting discussions. A.C. thanks NSERC, CIFAR and the McGill Centre for the Physics of Materials for support.
K.L.H. acknowledges the support from the Department of Energy in USA under the grant DE-FG02-08ER46541 
and is grateful to ENS Paris for the kind hospitality.

\appendix

\section{Counterterms and model renormalization}\label{counterterms}

The improper self-energy can be calculated to second order in $\varepsilon/T_K$
following Refs.~\cite{affleck1993,lehur2007}. The result is that the dependence
on the cutoff $D$ can be removed by adding the counterterm
\begin{equation}\label{counter1}
\begin{split}
H_{c,1} &=  - \frac{1}{2 \pi \nu T_K} \sum_{k,k',\sigma}
\,  \delta \alpha_1 \, (\varepsilon_k+\varepsilon_{k'}) \, 
: b^\dagger _{k\sigma} b_{k'\sigma} : \\[1mm]
\delta \alpha_1 &= - \alpha_1 
\frac{\phi_1}{T_K} \frac{6 D}{\pi} 
\ln \left( \frac{4}{3} \right)
\end{split}
\end{equation}
to the Hamiltonian $H_0 + H_{\rm int}$, Eqs.~\eqref{elastic},\eqref{interaction}. 
It corresponds to a renormalization of  $\alpha_1 \to \alpha_1 + \delta \alpha_1$.

We will now show that the second contribution that arises from 
Eq.~\eqref{curr1}, and that we have discarded in Sec.~\ref{inelastic},
produces a term linear in $D$ exactly cancelled by the 
counterterm~\eqref{counter1}.
 Using the identity $\Sigma^{++} (t) - \Sigma^{--} (t)
= {\rm sgn} (t) ( \Sigma^{-+}  (t) - \Sigma^{+-} (t) )$, it takes the form
\begin{equation}\label{deltai2}
\begin{split}
& \delta I^{(2)}_{\rm int}  = 
N (N-1)  (1-C^2) \frac{ e \pi}{2\, h}  
\left( \frac{\phi_1}{\pi \nu^2 T_K} \right)^2  {\cal S} 
  \times \\[2mm]
&  \int d t \, {\rm sgn} (t)
 (\Sigma^{-+} - \Sigma^{+-}) (t) \, i \pi \nu \Delta f (-t) + c.c.,
\end{split}
\end{equation}
where $\Delta f (t)$ is the time Fourier transform of
 $\Delta f (\varepsilon) = 
f_{\rm L} (\varepsilon) - f_{\rm R} (\varepsilon) $.
Inserting the Fourier transform of $F_0 \pm 1$,
\begin{equation}\label{intF0}
\begin{split}
&\int_{-D}^{D}  \frac{d \varepsilon}{i} \left( F_0 (\varepsilon) \pm 1 \right)
e^{-i \varepsilon t} = \frac{\pi T}{\sinh(\pi T t)} \times \\[1mm]
& ( 2  \cos^2 \theta e^{-i \mu_L t} + 2  \sin^2 \theta e^{-i \mu_R t})
- \frac{2 e^{\pm i D t}}{t} 
\end{split}
\end{equation}
in Eq.~\eqref{sigma} with Eq.~\eqref{Gbb}, it can be checked
that intermediate values of $t \sim 1/T,1/V$ give a vanishing
result for Eq.\eqref{deltai2} (integrand is odd in $t$).
Eq.\eqref{deltai2} is therefore dominated by small $t \sim 1/D$.
In that limit, $\Sigma^{+-} (t)\simeq \nu^3 (1-e^{i D t})^3/t^3$ and
 $\Sigma^{-+} (t) = (\Sigma^{+-} (t))^*$. $\Delta f (t) \simeq 
\Delta f (0) + t \Delta f' (0)$ is expanded to first order in $t$
since the zeroth order gives an odd integrand and a vanishing integral.
After some straigthforward algebraic manipulations,
we eventually find the result
\begin{equation}\label{deltai22}
\frac{\delta I^{(2)}_{\rm int}}{(1-C^2) N e  / h}
= - \sin 2 \delta_0 \, 
\int d \varepsilon \, \frac{\delta \alpha_1 \varepsilon}{T_K} 
[ f_{\rm L} (\varepsilon) - f_{\rm R} (\varepsilon) ],
\end{equation}
where we have used that 
\[
{\rm Im} \int_0^{+\infty} (d u/u^2) 
(1 - e^{i u})^3 = 3 \ln (3/4).
\]
When higher orders in $t$ are included in the expansion, 
corrections to Eq.~\eqref{deltai22}
are of order ${\cal O} (1/D)$ and completely vanish in the
universal limit $D \to + \infty$. In particular, the ${\cal O} (1)$
contribution vanishes by symmetry.
Finally, the counterterm~\eqref{counter1} gives an elastic contribution
to the current that can be computed along the lines 
of Sec.~\ref{elastic-current}. The result compensates exactly
Eq.~\eqref{deltai22}.

\begin{figure}
\includegraphics[width=5.5cm]{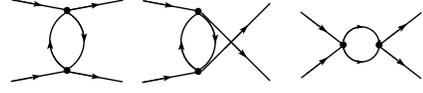}
\caption{\label{fvertex} $\phi_1^2$ corrections to the four-particle vertex.}
\end{figure}

A second counterterm is generated by vertex corrections.
In the spirit of the self-energy calculation, the singular 
contributions ({\it i.e.} depending on the cutoff $D$)
to the four-particle vertex are determined from the standard second order
diagrams shown in Fig.~\ref{fvertex}, and proportional to $\phi_1^2$.
This strong dependence on $D$ is removed by the counterterm
\begin{equation}\label{count2}
\begin{split}
H_{c,2} &=  \frac{\delta \phi_1}{\pi \nu^2 T_K} \sum_{\sigma < \sigma',\{ k_i \}} 
: b_{\sigma,k_1}^\dagger b_{\sigma,k_2}
b_{\sigma',k_3}^\dagger b_{\sigma',k_4}  :, \\[1mm]
\delta \phi_1 &= (N-2) \phi_1^2 \frac{D}{T_K} \frac{4}{\pi} \ln 2.
\end{split}
\end{equation}
To summarize, the perturbative calculation of observables
to second order in $\varepsilon/T_K$ ($\varepsilon$ is a typical energy,
$\mu_B B$, $k_B T$ or $e V$), from 
the full Hamiltonian $H_0 + H_{\rm int} + H_{c,1}+H_{c,2}$, 
 leads to finite and well-defined results
in the universal limit $D \to +\infty$.

\section{Finite temperature current} \label{finiteT}

We briefly outline how the current is calculated at finite temperature.
For the elastic part~\eqref{landauer} detailed in Sec.~\ref{elastic-current},
we merely need the Fourier transform of $\Delta f (\varepsilon) = 
f_{\rm L} (\varepsilon) - f_{\rm R} (\varepsilon) $,
\begin{equation}
\Delta f (t) = \left( e^{-i \mu_{\rm L} t} - e^{-i \mu_{\rm R} t} \right)
 \frac{i}{2 \pi} \frac{\pi T}{\sinh (\pi T t)}.
\end{equation}
The derivatives of $\Delta f (t)$, taken at $t=0$, give access to
the integrals with the corresponding powers of $\varepsilon$ 
in Eq.~\eqref{landauer}.

The inelastic part of the current is detailed in Sec.~\ref{inelastic} and
given by Eq.~\eqref{int-inel}.
Eq.~\eqref{int-inel} is evaluated at finite temperature by Fourier transform
to real time $t$. The time contour is then shifted by $i \eta$ in the complex
plane, with $\eta D \gg 1$ (but $T$, $e V \ll 1/\eta$) 
such as to suppress the dependence on the cutoff $D$
in Green's functions. From Eq.\eqref{intF0}, the result is (for $x=0$)
\[
{\cal G}_{bb}^{+-} (t) = - \nu \left( \cos^2 \theta e^{-i \mu_{\rm L} t} 
+ \sin^2 \theta e^{-i \mu_{\rm R} t} \right)
  \frac{\pi T}{\sinh (\pi T t)},
\]
with a similar expression for ${\cal G}_{bb}^{-+} (-t)$.
The intermediate integral result
\begin{equation}\label{intermediate}
\begin{split}
&\int_{-\infty+i\eta}^{+\infty+i\eta} d t \, [ {\cal G}_{bb}^{-+} (t)
{\cal G}_{bb}^{+-} (-t) {\cal G}_{bb}^{-+} (t)  - {\cal G}_{bb}^{+-} (t)
{\cal G}_{bb}^{-+} (-t) \\[1mm] 
& {\cal G}_{bb}^{+-} (t) ] i \pi \nu \Delta f (-t)
 = \pi \nu^4 \left[ \left( \frac{5}{12} - \frac{C^2}{4} \right) (e V)^2
+ \frac{2 (\pi T)^2}{3} \right] ,
\end{split}
\end{equation}
is used to derive the current interaction correction
\begin{equation}\label{intercorr2}
\begin{split}
 \frac{\delta I_{\rm int}}{(1-C^2) N e^2 V / h}
& = \cos 2 \delta_0 \,(N-1) \left( \frac{\phi_1}{ T_K} \right)^2 \times \\[1mm]
& \left[ \left( \frac{5}{12} - \frac{C^2}{4} \right) (e V)^2 
+ \frac{2 (\pi T)^2}{3} \right].
\end{split}
\end{equation}
The $t$-integral in Eq.~\eqref{intermediate} is obtained  by first
expanding the numerator in powers of $e^{\pm i \mu_{\rm L/R}}$.
Each term gives an integral. The standard method to evaluate such integrals
is to shift the integration contour by $-i/T$ in the complex
plane wich encloses the pole at $x=0$.

\section{Details on the interaction correction to the noise}\label{details}

We discuss in this Appendix the interaction corrections to the noise with two 
interaction vertices, {\it i.e.} the diagrams~\ref{diag}.b-f. Terms $\propto \phi_2,\chi_2$
in the interaction Hamiltonian~\eqref{interaction} are already
second order in $1/T_K$. Therefore only the term $\propto \phi_1$ is kept 
for the diagrams~\ref{diag}.b-f since the whole calculation goes up to second order
in $1/T_K$. 

In order to simplify the forthcoming expressions, let us define the following prefactor
\begin{equation}\label{prefac}
{\cal S}_P \equiv   \frac{h N (N-1) \sin^2 2 \theta}{\pi}
\left( \frac{e}{2 \nu h} \right)^2
\left( \frac{\phi_1}{\pi \nu^2 T_K} \right)^2.
\end{equation}
We start by considering the diagram~\ref{diag}.f where the self-energy bubble
is inserted in the top Green's function. 
Going to energy space and integrating over time $t$ in Eq.~\eqref{noise}, the 
corresponding contribution takes the form
\begin{equation}
\delta  S_{\rm int}^{f,1} = {\cal S}_P 
  \sum_{\eta_1,\eta_2} \eta_1
\eta_2  \int \frac{d \varepsilon}{2 \pi} \Sigma^{\eta_1,\eta_2} (\varepsilon) {\cal P}^{\eta_1,\eta_2}
(\varepsilon), 
\end{equation}
where $\Sigma^{\eta_1,\eta_2} (\varepsilon)$ is the self-energy part~\eqref{sigma} that 
already appeared in the calculation of the current. ${\cal P}^{\eta_1,\eta_2}$ is a notation for the
product of the three Green's functions (of the form ${\cal G}^{-\eta_1} {\cal G}^{\eta_2+} {\cal G}^{+-}$)
that enclose the self-energy in diagram~\ref{diag}.f.
Since the current operator~\eqref{current} has four different terms, this gives a sum of
$16$ terms for ${\cal P}^{\eta_1,\eta_2}$. Yet nine of these terms have no $\eta_{1/2}$
 dependence and vanish when summed over $\eta_{1/2}$. This is a consequence of the
causality identity~\eqref{causal}. Finally ${\cal P}^{\eta_1,\eta_2}$ reads
\widetext
\begin{equation}
\begin{split}
& {\cal P}^{\eta_1,\eta_2} (\varepsilon) = (- {\cal S}^*) {\cal G}_{ab}^{-\eta_1} (x,\varepsilon) {\cal G}_{bb}^{\eta_2+} (x, \varepsilon)
{\cal G}_{ab}^{+-} (-2 x,\varepsilon) 
+ (- {\cal S}^*)^2 {\cal G}_{ab}^{-\eta_1} (-x,\varepsilon) {\cal G}_{bb}^{\eta_2+} (x, \varepsilon)
{\cal G}_{ab}^{+-} (0,\varepsilon) \\[1mm]
& + (- {\cal S}) {\cal G}_{bb}^{-\eta_1} (-x,\varepsilon) {\cal G}_{ba}^{\eta_2+} (-x, \varepsilon)
{\cal G}_{ba}^{+-} (2 x,\varepsilon) 
+ (- {\cal S})^2 {\cal G}_{bb}^{-\eta_1} (-x,\varepsilon) {\cal G}_{ba}^{\eta_2+} (x, \varepsilon)
{\cal G}_{ba}^{+-} (0,\varepsilon) \\[1mm]
& + (- {\cal S}^*)(-{\cal S}) {\cal G}_{bb}^{-\eta_1} (-x,\varepsilon) {\cal G}_{bb}^{\eta_2+} (x, \varepsilon)
{\cal G}_{aa}^{+-} (0,\varepsilon) 
+ (- {\cal S}^*) {\cal G}_{bb}^{-\eta_1} (x,\varepsilon) {\cal G}_{bb}^{\eta_2+} (x, \varepsilon)
{\cal G}_{aa}^{+-} (-2 x,\varepsilon) \\[1mm]
& + (- {\cal S}) {\cal G}_{bb}^{-\eta_1} (-x,\varepsilon) {\cal G}_{bb}^{\eta_2+} (-x, \varepsilon)
{\cal G}_{aa}^{+-} (2 x,\varepsilon).
\end{split}
\end{equation}
The noise contribution with a bottom self-energy insertion gives a similar expression.
Green's functions are replaced by their expression~\eqref{nong} and the summation over $\eta_{1/2}$
is performed together with the causality identity~\eqref{causal}. In analogy with the current
calculation, two sorts of terms are obtained: (i) those including the combination $\Sigma^{++}
- \Sigma^{--}$ and (ii) those with $\Sigma^{+-}$ or $\Sigma^{-+}$. Type (i) terms are dominated
by energies on the order of the model cutoff $D$.  They are exactly cancelled by 
the counterterm~\eqref{counter1}. We therefore only keep type (ii) terms.
Combining top and bottom self-energy insertion diagrams, $\delta S_{\rm int}^f = \delta S_{\rm int}^{f,1}
+ \delta S_{\rm int}^{f,2}$, we find the contribution
\begin{equation}\label{deltaSf}
\begin{split}
\frac{\delta S_{\rm int}^f}{(1-C^2) 2 N e^2/h} & = - \frac{i}{2} \frac{(N-1) \phi_1^2}{\nu^3 T_K^2}
 \int \frac{d \varepsilon}{2 \pi} \Big( [ (1-C^2) ( \cos 4 \delta_0 -  \cos 2 \delta_0) [\Delta f (\varepsilon)]^2
+ 2 T_0 (F_0 (\varepsilon) \tilde{F}_0 (\varepsilon) -1) ] \times \\[1mm] & (\Sigma^{+-} (\varepsilon) - \Sigma^{-+}(\varepsilon))
- (\tilde{F}_0 (\varepsilon) -1) \Sigma^{+-} (\varepsilon) - (\tilde{F}_0 (\varepsilon) +1) \Sigma^{-+} (\varepsilon) \Big)
\end{split}
\end{equation}
At zero temperature, $F_0 (\varepsilon) \tilde{F}_0 (\varepsilon) -1 = - (1+C^2) \Delta f(\varepsilon)$ and
\begin{equation}
\int_{\mu_{\rm R}}^{\mu_{\rm L}} \frac{d \varepsilon}{2 \pi} \left( \Sigma^{+-} (\varepsilon) - \Sigma^{-+}(\varepsilon)
\right) = i \nu^3 \left( \frac{5}{12} - \frac{C^2}{4} \right) (e V)^3,
\end{equation} 
as we have shown in Sec.~\ref{inelastic} for the current. The last two terms in Eq.~\eqref{deltaSf}
involve the combination $4 J_1 (1 + C^4) + J_2 (1-C^4)$ with $J_{1/2}$ given Eqs.~\eqref{Js}.
Finally, we obtain for the noise correction~\eqref{deltaSf},
\begin{equation}\label{dsf}
\frac{\delta S_{\rm int}^f}{(1-C^2) 2 N e^3 |V| /h} = (N-1) \phi_1^2 \left( \frac{e V} {T_K} \right)^2
\left[ \frac{1}{4} - \frac{C^4}{12} + \left(\frac{5}{24} - \frac{C^2}{8} \right)
\left[ (1-C^2) ( \cos 4 \delta_0 - \cos 2 \delta_0 )- 2 T_0 (1+C^2) \right] \right].
\end{equation}
\endwidetext

\begin{figure}
\includegraphics[width=2.5cm]{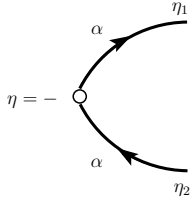}
\caption{\label{building} Building block appearing in diagrams~\ref{diag}.a-e
defined as $A^{\eta_1,\eta_2} (\varepsilon)$, see Eq.~\eqref{defA}. It is formed
by one current vertex on the branch $\eta =+$ or $-$ with two lines. 
The incoming line connects the branch $\eta_2$
to $\pm$ while the outgoing connects $\pm$ to $\eta_1$. 
}
\end{figure}

Next we turn to  the diagram~\ref{diag}.c with the particle-hole pair
 polarization bubble,
\begin{equation}\label{defpi}
\Pi^{\eta_1,\eta_2}  (t)  = \sum_{k_1,k_2} 
{\cal G}^{\eta_1,\eta_2}_{bb} (k_1,t) {\cal G}^{\eta_2,\eta_1}_{bb} (k_2,-t).
\end{equation}
Another {\it causality} identity, similar to Eq.~\eqref{causal}, also applies here.
For $t \ne 0$,
\begin{equation}\label{causal2}
\Pi^{++} (t) +  \Pi^{--} (t) = \Pi^{+-} (t) +  \Pi^{-+} (t).
\end{equation}
An elementary building block that appears in diagrams~\ref{diag}.a-e is shown
Fig~\ref{building}. 
It consists in one current vertex on the branch $\eta = +$ or $-$ 
supplemented by one incoming and one outgoing lines. In energy space, it reads 
\begin{equation}\label{defA}
A^{\eta_1,\eta_2} (\varepsilon) = (i \pi \nu)^2 \sin 2 \theta \Delta f (\varepsilon) 
\left[ 4 T_0 F_0(\varepsilon) + {\cal S} \eta_1 - {\cal S}^* \eta_2 \right],
\end{equation}
where the four terms of the current operator~\eqref{current} are included.
Using the definitions~\eqref{prefac},~\eqref{defpi},~\eqref{defA}, the noise term
due to diagram~\ref{diag}.c can be written
\begin{equation}
\begin{split}
\delta S_{\rm int}^c & = {\cal S}_P \sum_{\eta_1,\eta_2} \eta_1
\eta_2  \int \frac{d \varepsilon_1}{2 \pi} \int \frac{d \varepsilon_2}{2 \pi}
A^{\eta_1,\eta_2} (\varepsilon_1)  \\[1mm]
& \times \Pi^{\eta_1,\eta_2} (\varepsilon_1 - \varepsilon_2) 
A^{\eta_2,\eta_1} (\varepsilon_2).
\end{split}
\end{equation}
Following a now familiar pattern, there are terms with $\Pi^{++}-\Pi^{--}$
and others with $\Pi^{\pm,\mp}$. The former ones depend linearly on the
cutoff $D$ and are exactly cancelled by counterterms. This will be discussed
at the end of this Appendix.
\widetext
We are left with
\begin{equation}
\begin{split}
\frac{\delta S_{\rm int}^c}{(1-C^2) 2 N e^2/h} & =  \frac{\pi (N-1) (1-C^2) \phi_1^2}{2 \nu^2 T_K^2}
\int \frac{d \varepsilon_1}{2 \pi} \int \frac{d \varepsilon_2}{2 \pi}
\Delta f (\varepsilon_1) \Delta f (\varepsilon_2) \\[1mm]
& \left[ (\Pi^{+-}_{\varepsilon_1 - \varepsilon_2} +  \Pi^{-+}_{\varepsilon_1 - \varepsilon_2})  \cos 4 \delta_0 
+ 2 (\Pi^{+-}_{\varepsilon_1 - \varepsilon_2} -  \Pi^{-+}_{\varepsilon_1 - \varepsilon_2})   T_0
\, \cos 2 \delta_0 \, (F_0 (\varepsilon_1) - F_0 (\varepsilon_2)) \right].
\end{split}
\end{equation}
At zero temperature, the second term in the brackets gives a vanishing contribution.
Developping in terms of Fermi step functions, the integrals over energies can be
performed leading to
\begin{equation}\label{dsc}
\frac{\delta S_{\rm int}^c}{(1-C^2) 2 N e^3 |V| /h} = (N-1) \phi_1^2 \left( \frac{e V} {T_K} \right)^2
(1-C^2) \cos 4 \delta_0 \,  \left( \frac{1}{6} - \frac{C^2}{12} \right),
\end{equation}
corresponding to the combination $J_2 (1-C^2) + 4 J_1 C^2$.
\endwidetext
The diagram~\ref{diag}.d gives exactly the same contribution, $\delta S_{\rm int}^d
= \delta S_{\rm int}^c$. The calculation for the  diagram~\ref{diag}.e is quite
similar with the introduction of the particle-particle bubble,
\begin{equation}\label{defpit}
\tilde{\Pi}^{\eta_1,\eta_2}  (t)  = \sum_{k_1,k_2} 
{\cal G}^{\eta_1,\eta_2}_{bb} (k_1,t) {\cal G}^{\eta_1,\eta_2}_{bb} (k_2,t),
\end{equation}
that satisfies the same {\it causality} identity~\eqref{causal2}.
The noise term reads
\begin{equation}
\begin{split}
\delta S_{\rm int}^e & = {\cal S}_P \sum_{\eta_1,\eta_2} \eta_1
\eta_2  \int \frac{d \varepsilon_1}{2 \pi} \int \frac{d \varepsilon_2}{2 \pi}
A^{\eta_1,\eta_2} (\varepsilon_1)  \\[1mm]
& \times \tilde{\Pi}^{\eta_1,\eta_2} (\varepsilon_1 + \varepsilon_2) 
A^{\eta_1,\eta_2} (\varepsilon_2),
\end{split}
\end{equation}
leading to
\widetext
\begin{equation}
\begin{split}
\frac{\delta S_{\rm int}^e}{(1-C^2) 2 N e^2/h} & =  - \frac{\pi (N-1) (1-C^2) \phi_1^2}{2 \nu^2 T_K^2}
\int \frac{d \varepsilon_1}{2 \pi} \int \frac{d \varepsilon_2}{2 \pi}
\Delta f (\varepsilon_1) \Delta f (\varepsilon_2) \\[1mm]
& \left[ (\tilde{\Pi}^{+-}_{\varepsilon_1 + \varepsilon_2} +  \tilde{\Pi}^{-+}_{\varepsilon_1 + \varepsilon_2})  
+ 2 (\tilde{\Pi}^{+-}_{\varepsilon_1 + \varepsilon_2} -  \tilde{\Pi}^{-+}_{\varepsilon_1 + \varepsilon_2})   T_0
\, \cos 2 \delta_0 \, (F_0 (\varepsilon_1) + F_0 (\varepsilon_2)) \right],
\end{split}
\end{equation}
and
\begin{equation}\label{dse}
\frac{\delta S_{\rm int}^e}{(1-C^2) 2 N e^3 |V| /h} = (N-1) \phi_1^2 \left( \frac{e V} {T_K} \right)^2
(1-C^2)  \left( \frac{1}{6} + \frac{C^2}{12} + T_0 
\cos 2 \delta_0 \, C^2 \right),
\end{equation}
corresponding to the combination $J_2 (1+C^2) - 4 J_1 C^2 + 8 T_0 \cos 2 \delta_0 \, C^2 (J_2 - 2 J_1)$.
We finally consider the diagram~\ref{diag}.b that also involves the particle-hole bubble $\Pi$,
\begin{equation}
\delta S_{\rm int}^b  = - {\cal S}_P (N-1) \sum_{\eta_1,\eta_2} \eta_1
\eta_2  \int \frac{d \varepsilon_1}{2 \pi} \int \frac{d \varepsilon_2}{2 \pi}
 A^{\eta_1,\eta_1} (\varepsilon_1) \Pi^{\eta_1,\eta_2} (0) 
A^{\eta_2,\eta_2} (\varepsilon_2),
\end{equation}
with the definitions~\eqref{prefac},~\eqref{defpi} and~\eqref{defA}. Keeping only the terms
with $\Pi^{\pm,\mp}$, we obtain
\begin{equation}
\frac{\delta S_{\rm int}^b}{(1-C^2) 2 N e^2/h} =  (1-C^2) \sin^2 (2 \delta_0) \, \frac{\pi (N-1)^2 \phi_1^2}{\nu^2 T_K^2}
\left( \int \frac{d \varepsilon}{2 \pi}  \Delta f (\varepsilon)\right)^2 
(\Pi^{+-}_0 +  \Pi^{-+}_0).
\end{equation}
At zero temperature, $\Pi^{+-}_0 = \Pi^{-+}_0 = (\pi \nu^2 /2) (e V) (1-C^2)$ so that the noise 
correction for diagram~\ref{diag}.b finally reads
\begin{equation}\label{dsb}
\frac{\delta S_{\rm int}^b}{(1-C^2) 2 N e^3 |V| /h}  = \frac{(N-1)^2 \phi_1^2 \sin^2 (2 \delta_0)}{4}\,  
\left( \frac{e V} {T_K} \right)^2 (1-C^2)^2. 
\end{equation}

Before concluding this long Appendix, we briefly discuss the remaining terms resulting from the
$\Pi^{++} - \Pi^{--}$ and $\tilde{\Pi}^{++} - \tilde{\Pi}^{--}$ combinations. They lead to 
contributions that are linear in the cutoff $D$, with corrections scaling as ${\cal O} (1/D)$
and therefore vanishing in the universal limit. The calculation is straightforward and uses the same ingredients as in the 
Appendix~\ref{counterterms}, {\it i.e.} small $t$ dominate time integrals.
Therefore we can use $\Pi^{+-} (t) \simeq - \nu^3 (1-e^{i D t})^2/t^2$,
$\tilde{\Pi}^{+-} (t) \simeq - \Pi^{+-} (t)$  and
 $\Pi^{-+} (t) = (\Pi^{+-} (t))^*$, $\tilde{\Pi}^{-+} (t) = (\tilde{\Pi}^{+-} (t))^*$ in
those integrals. The final result reads
\begin{equation}
\frac{\delta S_{\rm int}^{(\propto D)}}{ (1-C^2)^2 2 N e^3 | V | /h} =  -
  \delta \phi_1 (N-1) \, C \sin 2 \delta_0 \, \sin^2 \delta_0 \frac{e V}{T_K},
\end{equation}
where we recall that $\delta \phi_1 = (N-2) \phi_1^2 \frac{D}{T_K} \frac{4}{\pi} \ln 2$.
This contribution is exactly cancelled by the  counterterm~\eqref{count2} included in the
diagram of Fig.~\ref{diag}.a.

\endwidetext

\newcommand{{{\PRB}}}{{{Phys. Rev. B}}}\newcommand{{{\PR}}}{{{Phys. Rev.}}}\newcommand{{{\PRA}}}{{{Phys. Rev. A}}}\newcommand{{{\PRL}}}{{{Phys. Rev. Lett}}}\newcommand{{{\NPB}}}{{{Nucl. Phys. B}}}\newcommand{{{\RMP}}}{{{Rev. Mod. Phys.}}}\newcommand{{{\ADV}}}{{{Adv. Phys.}}}\newcommand{{{\EPJB}}}{{{Eur. Phys. J. B}}}\newcommand{{{\EPJD}}}{{{Eur. Phys. J. D}}}\newcommand{{{\JPSJ}}}{{{J. Phys. Soc. Jpn.}}}\newcommand{{{\JLTP}}}{{{J. Low Temp. Phys.}}}\newcommand{{{\PTP}}}{{{Progr. Theoret. Phys.}}}

\end{document}